# Multi-Sensor Trajectory Reconstruction of the 24 April 2025 Alaska Fireball and Implications for Planetary Defense

L. T. Scamfer[1,2*], E. A. Silber[1], M.D. Fries[3], D. Vida[4], D. Šegon[5], P. Jenniskens[6,7], Y. Nishikawa[8], V. Sawal[1], T. A. Rector[9]

[1]Sandia National Laboratories, Albuquerque, NM 87123, USA
[2]Geophysical Institute, University of Alaska Fairbanks, Fairbanks, AK 99775, USA
[3]Astromaterials Research and Exploration Science Division, NASA Johnson Space Center, Houston, TX 77058, USA
[4]Department of Physics and Astronomy, Western University, London, ON, N6A 3K7, Canada
[5]Astronomical Society Istra Pula, Park Monte Zaro 2, HR-52100 Pula, Croatia
[6]SETI Institute, Mountain View, CA 94043, USA
[7]NASA Ames Research Center, Moffett Field, CA 94035, USA
[8]Osaka Kyoiku University, Division of Math, Sciences, and Information Technology in Education, Osaka, Japan
[9]Department of Physics & Astronomy, University of Alaska Anchorage, Anchorage, AK 99508, USA

*Corresponding author: ltscamf@sandia.gov

## Abstract

On 24 April 2025 at 18:30:57 UTC, a bright daytime fireball over Southcentral Alaska was detected by 37 seismic stations, 16 single infrasound sensors, and four infrasound arrays, yielding 30 ballistic and multiple fragmentation arrivals. The unprecedented density of seismoacoustic coverage enabled detailed reconstruction of the event using acoustic signals, with fragmentation source locations further guiding the identification of Doppler weather radar signatures of a meteorite fall. Incorporation of a radar-derived terminal point yielded a final trajectory solution, which agreed closely with an independent optical trajectory solution from video analysis. The reconstructed entry parameters from seismoacoustic analysis indicate a velocity of 25.3 km/s, an entry angle of 19°, and an energy release of ~38 t TNT equivalent. Assuming a chondritic composition, the pre-entry object diameter was ~0.7 m. Using orbital parameters from the optical solution, we estimate meteoroid composition as most likely a L-type ordinary chondrite. The event occurred in the sub-Arctic, where space-based optical systems face challenges in detection, demonstrating the critical role of dense ground-based seismoacoustic networks in characterizing high-latitude atmospheric entries. This uniquely well-recorded event demonstrates the capability of dense seismoacoustic networks to constrain bolide trajectories, energetics, and fragmentation, with radar and optical data providing critical confirmation and complementary perspectives. These results bridge the methodological gap between planetary-defense monitoring of natural impactors and space-traffic analyses of artificial reentries, illustrating how multi-sensor integration can deliver calibration-grade trajectories even for unpredicted events.

## Plain Language Summary

When large meteoroids or pieces of space debris enter Earth's atmosphere, they produce shock waves that travel long distances and can be recorded by special microphones (infrasound sensors) and seismometers. On April 24, 2025, a bright fireball crossed the daytime sky over Southcentral Alaska. The object, a large meteoroid about 0.7 meters in diameter (roughly the size of a car tire), entered Earth's atmosphere on a shallow path of about 19 degrees. Traveling more than 23 kilometers per second (>50,000 miles per hour) it released energy equivalent to ~38 tons of TNT before breaking apart. The entry generated acoustic waves which were detected by a dense network of instruments, some as far as 580 kilometers away. Doppler weather radar identified falling debris, and video cameras recorded three bright flashes caused by fragmentation. By combining these datasets, we reconstructed the meteoroid's trajectory with high precision and estimated its size. Because this event occurred in the sub-Arctic, where space-based sensors often have limited coverage, it demonstrates how ground-based monitoring networks can fill observational gaps. This event shows how multi-sensor data integration can effectively characterize fireball dynamics. These capabilities are important for understanding meteoroid and space-debris impacts and have applications in planetary defense.

**Key points:**

- A dense network of 57 seismoacoustic instruments recorded the 24 April 2025 Alaska fireball, enabling a trajectory reconstruction.
- Seismoacoustic analysis provided meteoroid trajectory and energy estimates.
- Seismoacoustic methods provide robust post-event characterization of bolides and space debris reentries, supporting planetary defense.

## 1. Introduction

Bright fireballs are rare yet scientifically valuable natural phenomena for studying the physics of high-energy atmospheric entries. However, due to their unpredictable and sporadic nature, only a small fraction are captured by dense, multi-modal geophysical networks. On 24 April 2025 at 18:30:57 UTC (~10:31 AM local time), a luminous daytime fireball over Southcentral Alaska became one of the most well-instrumented events ever recorded with seismoacoustic sensors. This event was detected by 37 seismic stations, 16 single infrasound sensors, and four infrasound arrays. Supplemented by optical imagery and Doppler radar evidence of a meteorite fall, this event presents a rare opportunity to resolve fireball trajectory, fragmentation dynamics, and energy deposition.

Asteroid entries can pose significant risks to human populations and infrastructure (Boslough et al., 2015; Boslough & Crawford, 2008; Chapman, 2008; Cohen et al., 2024 Trigo-Rodríguez, 2022). The 2013 Chelyabinsk superbolide, for example, injured more than 1,600 people, damaged over 7,000 structures (Popova et al., 2013), and released ~30× the yield of the 1945 Hiroshima nuclear detonation (Brown et al., 2013). Detecting and characterizing such events is instrumental for constraining the flux of near-Earth objects and improving planetary defense readiness (Bailey et al., 2025; Bender et al., 1995; Landis & Johnson, 2019; Mainzer, 2017; Silber, 2024b). Although many significant bolides are detected by U.S. Government space-based sensors and reported to the National Aeronautics and Space Administration's (NASA) Center for Near-Earth Object Studies, ground-based all-sky cameras (e.g., Devillepoix et al., 2019; Devillepoix et al., 2020; Ehlert & Erskine, 2020) and seismoacoustic networks play a vital role in providing coverage where space-based optical systems are unavailable or obstructed.

These same measurement principles apply to space debris reentries, which generate high temperature flow fields behind hypersonic shock fronts and frequent fragmentation episodes analogous to natural bolides. Planned reentries such as the Origins, Spectral Interpretation, Resource Identification, and Security-Regolith Explorer (OSIRIS-Rex) Sample Return Capsule (Silber et al., 2024a; Silber & Bowman, 2025) and Hayabusa2 (Nishikawa et al., 2022) have demonstrated the use of infrasound for validating propagation models, atmospheric effects, and network performance (Nishikawa et al., 2022; Sansom et al., 2022; Silber et al., 2025). In both planetary defense and space traffic management contexts, the ability to perform rapid, forensic post-event reconstruction is critical for situational awareness and public communication.

Meteoroids enter the Earth's atmosphere at velocities of ~11–73 km/s (Ceplecha et al., 1998). As an object penetrates denser atmospheric layers and the surrounding flow transitions into the continuum regime (see Silber et al., 2018 for discussion), shock waves are generated via two primary mechanisms (e.g., Wilson et al., 2025). The first is the cylindrical line source associated with the hypersonic shock front forming along the luminous path. This shock originates within the strongly non-linear near field bounded by the blast radius ($R_0$) (ReVelle, 1976). As the disturbance propagates outward, it transitions to a weak shock beyond ~$10R_0$, and ultimately evolves into linear acoustic waves. The shock geometry is governed by the Mach cone, with half-angle $\theta_M = \arcsin(c/v)$, where $c$ is the local adiabatic sound speed and $v$ is the meteoroid velocity. For hypersonic entries ($v \gg c$), $\theta_M$ is small, and the shock envelope is well approximated as a cylindrical line source aligned with the trajectory (ReVelle, 1976). This geometry enables directional resolution of ballistic arrivals using infrasound arrays (Evers & Haak, 2003; Le Pichon et al., 2002; Silber & Brown, 2014). The second mechanism is fragmentation, either continuous or discrete, which releases energy in localized volumes and is commonly approximated as a quasi-spherical point source (McFadden et al., 2021; Pilger et al., 2020; Trigo-Rodríguez et al., 2021). Both source types were observed during the Alaska event, with stations recording impulsive ballistic N-waves (Dumond et al., 1946) and lower-frequency arrivals associated with discrete fragmentation episodes.

Acoustic waves below ~20 Hz, or infrasound, can propagate over many hundreds of kilometers with minimal attenuation (e. g., Hedlin et al., 2012). Therefore, infrasound is widely used to study energetic atmospheric phenomena, including volcanic eruptions (Gheri et al., 2025; Watson et al., 2022), large explosions (Berg & Poppeliers, 2022; Golden et al., 2012; Park et al., 2018), mass movements (Johnson et al., 2023; Sovilla et al., 2025), earthquakes (Anderson et al., 2023; Macpherson et al., 2023), and bolides (Arrowsmith et al., 2005; Belli et al., 2021; Brown et al., 2002). Infrasound arrays, such as those in the International Monitoring System (IMS) operated by the Preparatory Commission of the Comprehensive Nuclear-Test-Ban Treaty Organization (Christie & Campus, 2009), use beamforming to determine the back-azimuth and trace velocity of incoming waves (Bishop et al., 2020; Cansi & Le Pichon, 2008). One such IMS station, I53US in Fairbanks, Alaska, provided directional data in this study.

Seismic detections of fireballs arise primarily through air-to-ground coupling, whereby acoustic waves induce measurable ground motion (Edwards et al., 2008; Langston, 2004). Because seismic networks are often denser than dedicated infrasound networks, they are increasingly leveraged to augment bolide analyses (Ishihara et al., 2004; Pujol et al., 2006; Hedlin et al., 2010; Kalenda et al., 2014; Hicks et al., 2023). Rarely, Rayleigh waves can be excited by intense airbursts (Tauzin et al., 2013) or by meteorite impacts (Le Pichon et al., 2008).

The Alaska fireball occurred in an ideal observational setting, with extensive co-located seismic and acoustic instrumentation originally deployed for earthquake and volcanic monitoring. Here we integrate seismic, infrasound, optical, and radar datasets to reconstruct the fireball trajectory. Unlike most bolide studies, where optical measurements serve as the primary constraint, we invert this paradigm by first using infrasound and seismic signals to refine the trajectory, then applying Doppler weather radar guided by infrasound-derived fragmentation points to identify the terminal location and finally validating the solution with independent optical observations. We quantify the trajectory, velocity, and energy, identify fragmentation episodes, and evaluate the complementary roles of seismic and infrasound methods in bolide detection. While the exceptional density of seismoacoustic coverage in the sub-Arctic provided a rare opportunity, the methodological framework demonstrated here is broadly applicable to bolide events in any geographic setting, making this case a benchmark for understanding the geophysical signatures of high-energy meteoroid entries.

## 2. Methods

### 2.1 Data Sources and Network Description

The 24 April 2025 Southcentral Alaska fireball was recorded by a dense regional seismoacoustic network consisting of 37 broadband seismic stations, 16 single infrasound sensors, and four infrasound arrays (Figure 1). Seismic and infrasound stations are operated by the Alaska Earthquake Center (AEC), the Alaska Volcano Observatory (AVO), the U.S. National Tsunami Warning Center, and the IMS. Being part of permanent monitoring networks, the geophysical instruments are broadband and observatory-grade. Infrasound sampling rates range from 20–100Hz and seismometers sampled from 40–50Hz. Detailed information on instrumentation can be found where seismoacoustic data is hosted, at the Incorporated Research Institutions for Seismology (IRIS) EarthScope Data Management Center (https://service.iris.edu).

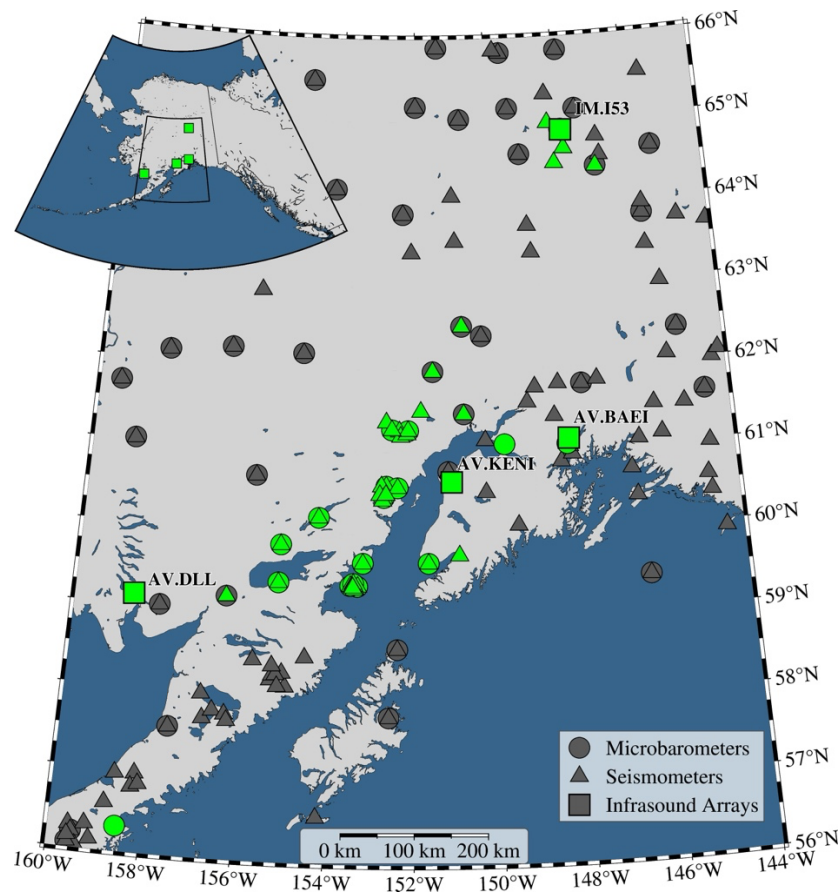

**Figure 1:** Detection map and regional seismoacoustic network in Southcentral and Interior Alaska. Only stations with available data during the event are shown. Green stations denote sensors with detections.

## 2.2 Initial Event Characterization and Search Strategy

A preliminary ground track trajectory for the 24 April 2025 Southcentral Alaska fireball was obtained from the International Meteor Organization (IMO) Fireball Events database (Event 2280-2025) to kickstart our solution. This trajectory had limited accuracy as it was derived from a few public visual observation reports. To improve the initial geometry, we used the first-arrival times from nearby seismoacoustic stations and time-dependent changes in signal back-azimuth from regional infrasound arrays to generate a preliminary seismoacoustic-based trajectory estimate. This geometry was then used to guide targeted searches for meteorite signatures in Doppler weather radar data. A publicly posted video of the event, recorded by Mr. Steve Friend and shared via the IMO website and with us, was also examined. The video provided accurate timing of the luminous phase and showed three distinct fragmentation episodes near the end of the trajectory. Additionally, it provided line-of-sight information pointing towards the luminous path. Using these timing markers, a high-confidence event window was established, and a signal arrival probability window was constructed by accounting for expected acoustic travel times from multiple atmospheric propagation paths (Silber, 2024a). Seismic data were examined within a 600 km radius and infrasound data within a 1,000 km radius of the revised trajectory midpoint. These distances were chosen to encompass near-field ballistic arrivals and distant fragmentation signals while remaining within reasonable propagation ranges that preserve signal characteristics (Sutherland & Bass, 2004).

## 2.3 Waveform Preprocessing, Signal Detection and Classification

All waveform data were detrended, demeaned, and corrected for instrument response to yield ground displacement (seismic) or pressure (infrasound) time series. To first identify relevant signals and pick arrival times, infrasound data were filtered using a 4th-order Butterworth bandpass between 0.3–5 Hz, matching the dominant frequency range of bolide infrasound (Ens et al., 2012). Seismic data were filtered between 2–8 Hz to optimize the signal-to-noise ratio for air-to-ground coupled waves (Wills et al., 2022).

Candidate detections were identified by visual inspection and confirmed if they exhibited a transient onset distinct from background noise, a signal-to-noise ratio at least three times greater than the pre-event level, and timing consistent with acoustic propagation speeds between 280 and 340 m/s. For infrasound arrays, additional confirmation required that back-azimuth estimates intersected the preliminary trajectory within ±10°, and were correlated across the array. For each infrasound array, waveforms were filtered using a passband optimized to the local noise environment. Passbands were selected based on frequency-domain signal-to-noise measurements to preserve the dominant spectral energy of the fragmentation signal while suppressing background noise. Array processing employed a least-trimmed squares beamforming algorithm (Bishop et al., 2020), which is robust against outlier sensors and improves stability in low signal-to-noise conditions. Confirmed arrivals were classified as ballistic if they displayed impulsive N-waveforms (DuMond et al., 1946) and short durations consistent with a cylindrical line source (ReVelle, 1976) or as fragmentation if they were emergent, longer duration, and consistent with quasi-spherical point sources (McFadden et al., 2021; Pilger et al., 2020; Silber & Brown, 2014; Wilson et al., 2025). The dominant signal period was measured using well-established 4-zero-crossing method for bolides applied to the filtered waveform (ReVelle 1997; Ens et al., 2012; Silber et al., 2025). For completeness, we also applied a 2-zero-crossing approach, adopted by Silber & Bowman (2025) for the analysis of an artificial object re-entry, and found it to be within ±10% of the 4-crossing method. Hereafter, any use or mention of signal period will refer to measurements using the 4-crossing method. The maximum and peak-to-peak amplitudes were also measured at the highest-pressure portion of the signal (Ens et al., 2012; Silber et al., 2025).

## 2.4 Energy Estimation

Energy yield ($E$) in kilotons TNT (1 kt TNT = 4.184E12 Joules) was estimated using bolide-specific period-yield relationships (ReVelle, 1997) of the form:

$$\log(E) = A \log(P) + B,$$

where $P$ is the dominant period in seconds, and $A$ and $B$ are empirically derived coefficients, with values varying across literature (Table 1 in Silber et al., 2025). Here we detected both ballistic and fragmentation arrivals, which reflect distinct modes of energy release. To explore this, we computed average signal period separately for arrivals determined as ballistic, and for those attributed to fragmentation. For our overall energy estimates, to reduce variability arising from stations sampling different portions of the trail, we used the recently refined coefficients applicable to the averaged period across all detections ($A$ =3.71 and $B$ =-2.07) (Silber et al., 2025). This approach provides the most stable estimate because it resolves differences between ballistic and fragmentation arrivals and captures the underlying source characteristics. The resulting energy was converted to mass and diameter assuming the optically-derived velocity (Section 3.6) and a chondritic bulk density of 3.36 ± 0.16 g/cm$^3$ (Consolmagno et al., 2008).

## 2.5 Doppler Weather Radar Analysis

Once the preliminary seismoacoustic-based trajectory was determined, it was used to narrow the spatial search area for meteorite fall signatures in Doppler weather radar data. Weather radar data is an established method for detection of meteorite falls and facilitating their rapid recovery (Fries & Fries, 2010). The radars themselves are designed for interrogating weather phenomena at altitudes typically below ~5 km by measuring radar reflectivity, velocity via Doppler shift, and three modes based on dual-polarized radar pulses. As a result, the radars do not typically detect the optically bright fireball itself, but rather the cross-sectional area of the surviving meteorite cloud during free-fall to the ground (or "dark flight"). The location of a strewn field of meteorites was modeled using the Jörmungandr dark flight model (Fries, 2023) version 2, which assumes the composition of an ordinary chondrite and accounts for atmospheric winds using radiosonde data collected by a balloon launched the same day at 12:00 UTC from Anchorage, AK. Weather radars are the only technique that can characterize meteorites during dark flight which gives them a unique role in meteorite fall analysis.

Radar imagery was obtained from the Next Generation Weather Radar (NEXRAD) national weather radar network operated by National Oceanic and Atmospheric Administration (NOAA), which distributes radar data through an openly accessible internet site. The NEXRAD network consists of 162 WSR-88D (Weather Surveillance Radar) pulsed-Doppler radars operating at a 10 cm wavelength, providing nearly complete coverage of the contiguous 48 states and partial coverage of Alaska. NEXRAD radars operate continuously, scanning through a pre-programmed series of tilt angles. When the series, or Volume Coverage Pattern (VCP) is complete, the radar formats the data into a package that is conveyed to a publicly available internet server where it is immediately available for download.

Luckily this event occurred within the interaction volumes of the PAHG (Anchorage, AK) at a distance of ~150 km and nominally within range of PAKC (King Salmon, AK) at a distance of ~280 km. Radar signatures of falling meteorites occur in PAHG data but not in PAKG data owing to the greater range. Falling meteorites are size-sorted in their ~25 km fall to Earth with 1 kg meteorites reaching the ground in approximately 90 s while 1g meteorites take about 10 minutes to cover the same distance. We use these differences in fall time and the distances between multiple radar signature to estimate the expected mass of meteorite that reach the surface (Fries, 2023). Radar data from all nearby radars were examined at the time and place of the event, searching for reflectivity signatures following a general high- to low-altitude trend, lasting ~10 minutes after the event, and following a trend of increasing reflectivity as sequentially more numerous, smaller meteorites fall into the radar's interaction volume. Doppler data were examined for a "candy striping" signature that is emblematic of meteorite falls. This signature appears within the first few minutes of the first appearance of meteorites and is derived from intense, short-range turbulence caused by the vertical passage of supersonic or transonic meteorites. The strongest radar signature, corrected for wind drift using the dark-flight model, was used to anchor a terminal location for the fall, which in turn was adopted as a fixed reference point for refining the seismoacoustic trajectory inversion.

## 2.6 Trajectory Reconstruction

### 2.6.1 Trajectory Inversion from Seismoacoustic Travel Times

The radar-derived meteorite fall location served as the terminal point in a five-dimensional parameter search over trajectory length, azimuth, entry angle, terminal altitude, and atmospheric celerity (for travel time calculations). We note that representing infrasound propagation using a fixed celerity implicitly assumes a constant effective sound speed along the propagation path and does not account for wind-induced path bending. This simplification is appropriate for the trajectory inversion because relative arrival-time differences dominate the solution geometry, and the regional station distances considered here limit the sensitivity to higher-order atmospheric structure. For the initial search, we used a coarse spacing for all parameters (e.g. 10 km spacing for terminal altitude, 5° spacing for azimuth, etc.), and once a coarse solution was derived we increased the resolution of the parameter search to obtain a more accurate solution. The final parameter spacing was as follows: 0.5 km for length, 0.01° for azimuth, 0.05° for entry angle, 0.1 km for terminal altitude, and 1 m/s for atmospheric celerity. The search over varying celerities was successfully applied to the Oregon Bolide (Walker et al., 2010) and is the approach we use here. To realistically compute the observed travel times to all stations, we first defined a time vector spanning the full duration of the fireball's flight, as constrained by video observations. The start time (18:30:57 UTC) and end time (18:31:04 UTC) corresponded to the first and last frames in which the fireball was visible. This time vector allowed us to assign a signal origin time for each station based on the point along the trial trajectory that minimized the three-dimensional (3D) source–receiver distance.

For each trial trajectory, predicted acoustic travel times were calculated for all stations using the shortest 3D distance from the trajectory to the station and each trial celerity. The misfit between predicted and observed arrival times was quantified as the sum of absolute differences across all stations, and the optimal trajectory was determined by minimizing this misfit. This approach is similar to techniques applied in previous bolide trajectory inversions (Che et al., 2016; Olivieri et al., 2023; Vida, 2023), but explicitly accounts for the variations in source–receiver geometry and origin time along the luminous path.

To avoid introducing bias in the misfit function, stations with co-located infrasound and seismic sensors contributed only a single arrival time, chosen from the sensor with the clearest ballistic arrival. Only stations that recorded ballistic arrivals, consistent with direct shock wave propagation from the meteoroid's hypersonic flight, were included in the inversion. Stations that detected only fragmentation-related signals, or whose arrival times were ambiguous, were excluded from the analysis.

As a validation of (and comparison to) our initial inversion method in Section 2.6.1 we tried an established, but alternate inversion scheme using the approach described in Nishikawa et al., (2022). The full methods of this inversion, which does not employ radar anchoring, can be found in Text S1 in the Supporting Information, and where we also present the results for comparison.

### *2.6.2 Independent 3D Point from the KENI Array*

While three arrays (BAEI, I53US, and DLL) captured exclusively fragmentation-generated infrasound, the KENI array recorded the decayed ballistic shock wave. This arrival, a product of the cylindrical line source radiating from the fireball's hypersonic trajectory, was particularly valuable for deriving an independent 3D point along the path. Array processing using a least-trimmed squares algorithm (Bishop et al., 2020) produced back-azimuth and trace velocity estimates, from which the incidence angle of the wavefront was determined. A parameter space of trial horizontal ranges (30–80 km) and altitudes (50–90 km), spaced at 0.5 km intervals along the back-azimuth from KENI, was searched. For each trial location, infrasound propagation modeling was performed with infraGA (Blom, 2019) using realistic atmospheric specifications from Ground-2-Space (G2S) (Drob et al., 2003; Hetzer, 2024). Rays were launched toward the array in 0.25° increments, and both theoretical travel time and incidence angle were computed. A normalized misfit function combining the differences between observed and modeled travel times and incidence angles was used to identify the best-fitting source location for each trial altitude $h$ and range $r$:

$$misfit(h,r) = \frac{|tt_{theo}(h,r) - tt_{obs}|}{tt_{obs}} + \frac{|\theta_{theo}(h,r) - \theta_{obs}|}{\theta_{obs}}, \qquad (2)$$

where $\theta$ is the incidence angle in degrees and $tt$ is the travel time in seconds. The altitude and range that minimizes this misfit function is the estimated source location. This independently derived point was then compared with the travel-time inversion trajectory to assess consistency.

## 2.7 Fragmentation Episode Geolocation

To corroborate our fireball trajectory estimates, we examined infrasound array recordings for signatures of discrete fragmentation episodes. These episodes were identified as temporally separated acoustic energy packets exhibiting stable back-azimuths over their duration, consistent with localized impulsive sources along the flight path. Within each filtered waveform, discrete acoustic "packets" were identified as candidates for fragmentation events. At the point of maximum amplitude for each packet, we measured the back-azimuth and back-azimuth uncertainty. These direction-of-arrival estimates were then used to construct great-circle paths from each array toward the inferred source region. Intersections of back-azimuths from multiple arrays provided preliminary two-dimensional geographic source locations.

## 2.8 Optical

The general public took several casual video recordings of the fireball, and we were able to compute an independent trajectory solution solely based on optical data. Two fixed security cameras operated by Mr. Steve Friend located 12 km West of Anchor Point captured the fireball from the beginning down to a height of ~33.6 km (Figure S1). We obtained calibration images that contained numerous stars for one camera (Camera 1), while the other (Camera 2) showed a noticeable shift, which prevented a quality calibration. For Camera 1, we achieved an astrometric fit accuracy of 3.7 arcmin using the SkyFit2 software (Vida et al., 2021a). As the fireball occurred during the day, nighttime video data from January 30, 2025 was used. We confirmed that there was no appreciable camera movement over the months until April by comparing the pixel positions of static objects in the image (a roof of an adjacent building, an antenna, and five pine trees). The astrometric calibration was done on 52 directly visible stars and the radial distortion of the lens was modeled using a 7th-order polynomial with odd terms only.

The other two videos were captured by a dash camera at Aurora Elementary School (https://www.alaskasnewssource.com/video/2025/04/24/video-fireball-streaks-across-alaska-skies/) and a dash camera of a car stopped at the intersection of Boniface Parkway and East Tudor Road in Anchorage. We were unsuccessful in obtaining calibration images for the school camera, but we were able to calibrate the dash camera video with nighttime images from 30 September 2025, following the approach described in Vida et al. (2021b). For accuracy relative to the original dashcam video, calibration images were taken from a tripod 1.2 m high in the left-turn lane of the street with a Canon 5D Mk3 paired with a Sigma 8mm fisheye lens. The trajectory and the associated uncertainties were computed using the method of Vida et al. (2020). Unfortunately, the fireball was not optically observed in its entirety due to local obstructions.

To compare the optically derived trajectory to the seismoacoustic trajectory, we first truncated the seismoacoustic trajectory by the minimum and maximum longitude and latitude of the optical solution. Next, we ensure that both of the trajectory vectors have an equal length, so we could do a point-by-point comparison of the 3D distance between the two.

## 3. Results

### 3.1 Seismoacoustic Detections and Signal Measurements

Approximately 155 seconds after the first optical signature of the fireball, an air-to-ground coupled wave was recorded at seismic station RDJH. This marked the onset of a sequence of seismoacoustic detections that, over the next 30 minutes, were observed by 37 seismic stations, 16 single infrasound sensors, and four infrasound arrays distributed across Southcentral Alaska (Figure 1). These detections span source–receiver distances from ~48 to 584 km along and off the reconstructed trajectory. This event produced one of the most spatially and temporally extensive seismoacoustic datasets ever recorded for a fireball, enabling detailed analyses of both propagation and source characteristics.

Arrival times and waveform morphologies were consistent with direct infrasound arrivals and air-to-ground coupling of infrasound. Of the 37 seismoacoustic stations, 30 recorded ballistic arrivals associated with the hypersonic shock front, while the remainder detected only fragmentation-generated signals. Maximum infrasound amplitudes across the network ranged from 0.08 to 14.3 Pa (Table S1, S3), with ballistic arrivals generally producing higher amplitudes at comparable distances than fragmentation episodes. Representative acoustic waveforms for both arrival types are shown in Figure S2. While generally acoustic amplitudes decay as distance from the trajectory increases, some amplitudes showed wide variability over local (<20 km) regions. For example, three infrasound sensors in the Alaska Volcano

Observatory's Redoubt volcano network recorded maximum amplitudes ranging from 2.4 to 14.3 Pa despite the maximum inter-station distance being only 15.5 km.

An interesting observation in the air-to-ground coupled seismic data is the potential presence of precursory seismic waves (Figure S3). Observations of these seismic waves are rare and precede the acoustic signal recorded at a station and occur under specific coupling scenarios which allow for a wave coupled at a separate location to arrive first (Edwards et al., 2008).

Notably, three stations (KENI, RC01, and HOM) clearly recorded both ballistic and fragmentation arrivals, allowing direct comparison of their spectral and temporal characteristics at a single receiver site. Other stations likely recorded fragmentation-related energy, but the corresponding arrivals could not be reliably distinguished from propagation-induced coda or noise and are therefore not individually interpreted. Infrasound array detections were broadly consistent with back-azimuth estimates toward the reconstructed trajectory. The KENI array uniquely recorded a ballistic N-wave in addition to fragmentation episodes, while the other three arrays (BAEI, I53US, and DLL) detected only fragmentation arrivals (Figure 2).

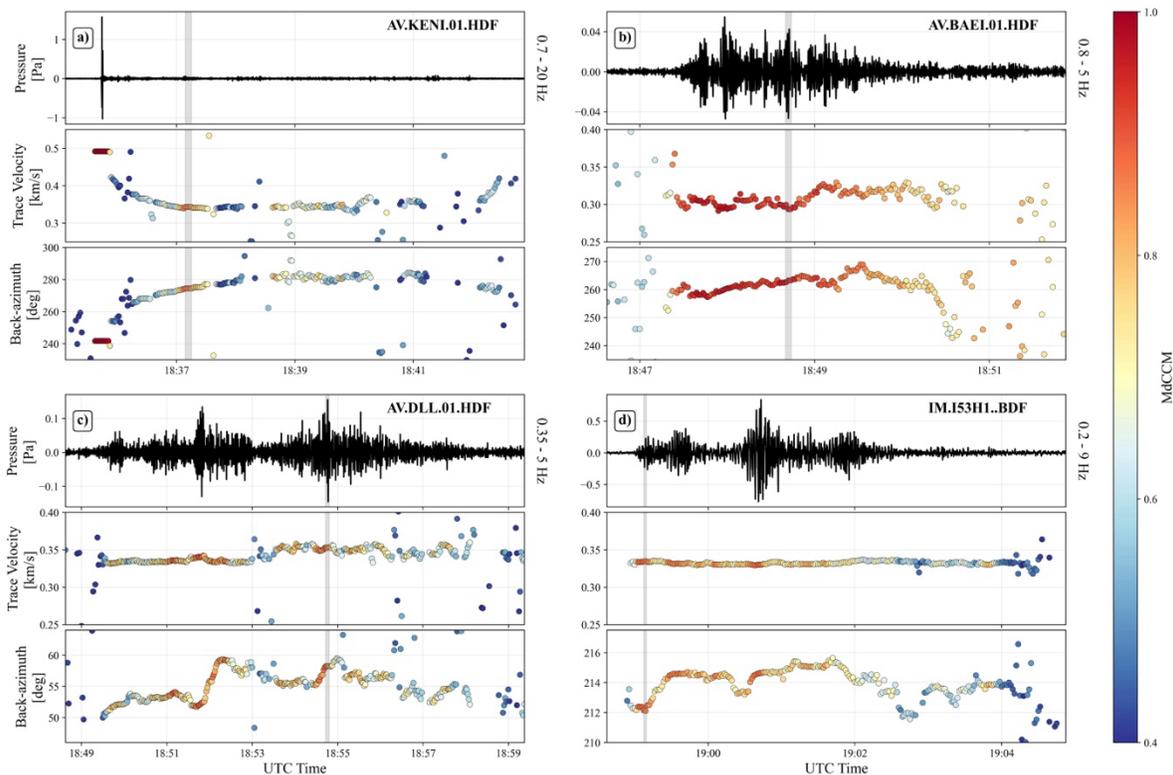

**Figure 2:** Array processing results from all 4 infrasound arrays that received a fireball signal. Each panel is a separate array, noted by the miniSEED Network, Station, Location and Channel codes. The grey highlighted section represents packet back-azimuth values seen

in Figure 4b to identify a fragmentation episode. Filter bands for each array are specified to the right of the waveforms. Circle colors represent the median cross-correlation maxima.

Fragmentation detections frequently appeared as discrete acoustic energy packets in the array data, with back-azimuth shifts that aligned with multiple source points along the trajectory. This pattern is consistent with temporally separated, spatially distributed fragmentation episodes occurring over tens of kilometers of the flight path. Such directional variations, when cross-referenced with trajectory reconstructions, provided an independent means of constraining fragmentation source locations (Section 3.3). The complete set of signal detections and measured parameters is summarized in Tables S1-S3.

## 3.2 Doppler Weather Radar Signatures

Three radar signatures were recorded for this event that are consistent with falling meteorites (Figure 3). Meteorite fall signatures were identified in data from the PAHG radar (Anchorage, AK) but not in PAKC (King Salmon, AK) data, likely due to the greater range from the fall site. The first signature, at 5.14° elevation in the 18:25 UTC dataset (which is named after the timing of the initial radar sweep in the dataset), occurred at 17,950 m above mean sea level (AMSL) 72 s after the fireball terminus, with an estimated mass of 17.76 g. The second, at 6.45° elevation in the same dataset, was observed at 15,528 m AMSL 91 s after the terminus, with an estimated mass of 0.27 g. The third signature appeared at 1.32° elevation in the 18:32 UTC dataset at 4,891 m AMSL 317 s after the terminus, with an estimated mass of 1.39 g. The three radar signatures were recorded at 18:32:10.9, 18:32:29.9, and 18:36:15.9 UTC, respectively. The gram-scale masses (17.76 g, 0.27 g, and 1.39 g) associated with these signatures represent radar reflectivity-equivalent particle masses derived from individual reflectivity bins and do not correspond directly to surviving meteorite fragment masses.

The first two signatures displayed pronounced Doppler velocity "candy striping," (Figure S4) indicative of turbulent wakes from meteorites in the 10s of grams mass range. Modeling with the Jörmungandr 2 dark flight model (Fries, 2023) placed the centroid of the strewn field at 60.9964° N, 154.0359° W (Figure 3). This centroid reflects the observable mass-weighted debris distribution after dark flight. Because fragment descent depends on assumed mass, starting altitude, and wind structure, dark-flight modeling yields a distributed impact field rather than a unique terminal coordinate. The radar-derived centroid was therefore adopted as the terminal constraint in the seismoacoustic trajectory inversion (Section 3.4), as it represents the only directly observable ground-impact indicator and appropriately reflects inherent dark-flight uncertainty. Although no field search has yet been conducted to recover meteorites, recovery may be possible in the future. The predicted strewn field lies in rugged terrain, and access limitations have so far precluded an on-site investigation.

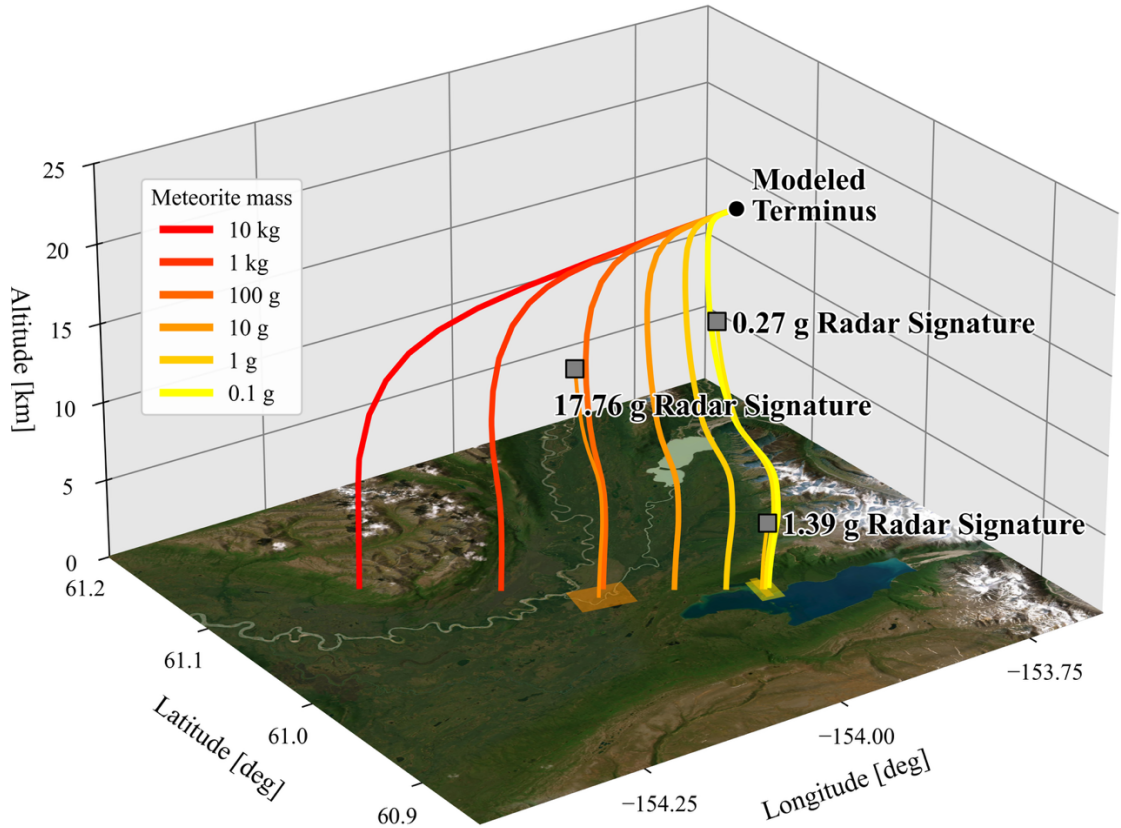

**Figure 3:** Jörmungandr 2 dark flight modeling using radar signatures of falling meteorites as detected by the PAHG radar (Anchorage, AK) from the NEXRAD national weather radar network. The grey squares represent three signatures detected in successive radar sweeps. Flight paths for a wide range of meteorite masses were calculated using the Jörmungandr dark flight model to produce the colored lines that descend towards the ground. Individual paths for meteorites falling from the radar signatures are also shown, along with their respective strewn field.

### 3.3 Fragmentation Episode Geolocation

The video timing of three distinct luminous fragmentation episodes near the end of the trajectory provides an independent visual reference for fragmentation. Infrasound analysis identified multiple acoustic fragmentation sources along the trajectory, one of which aligns closely in timing with the second luminous episode. Back-azimuth intersections from all four arrays (Figure 2) converged on a primary fragmentation location at 60.7028° N, 150.0598° W (Figure 4b). Additional fragmentation episodes were detected, but their localization was less constrained due to limited cross-array correlation.

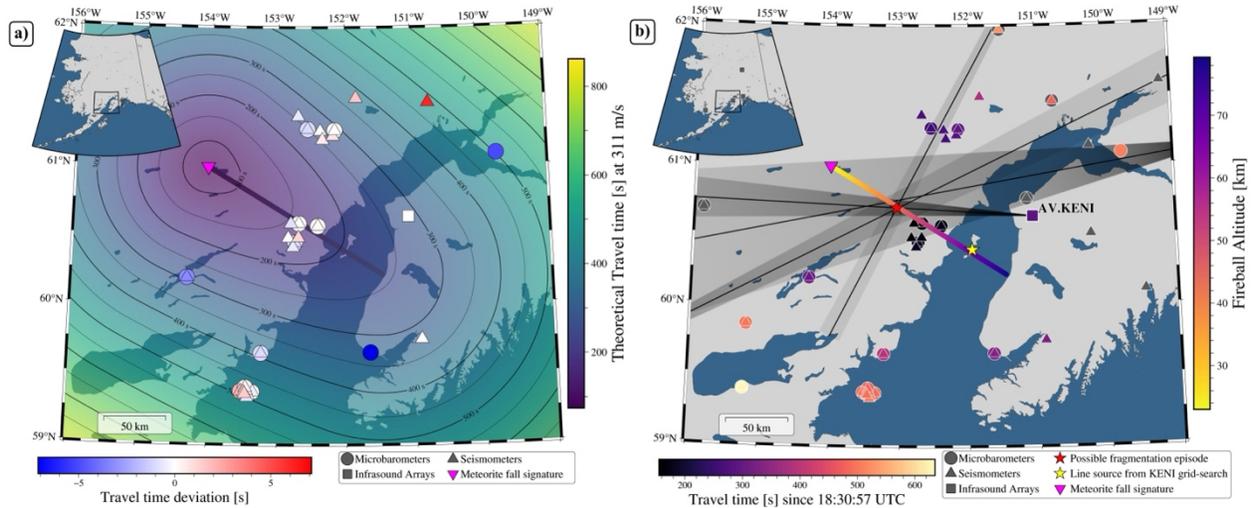

**Figure 4:** a) Map view of theoretical travel times to the ground surface from the best-fitting trajectory using the single celerity (311 m/s), straight-ray approximation. The thick grey line in the background is the trajectory projected to 2D. Seismoacoustic stations are colored by the deviation (in seconds) between theoretical and observed travel times. Only stations that recorded ballistic signals are shown. b) Map of regional geophysical instrumentation around the fireball. The colored thick line is the seismoacoustic-derived fireball trajectory. Seismic and infrasound sensors with detections are colored by travel time, otherwise they are left grey. Black lines indicate infrasound array back-azimuth estimates for acoustic energy packets (see Figure 2) that converge on an inferred fragmentation event, marked by a red star. The shaded regions are the 90% confidence intervals. The upside-down triangle (60.9964° N, 154.0359° W) marks the meteorite fall centroid (Figure 3), and the yellow star indicates the best-fitting point along trajectory using the ballistic arrival at the KENI array (see Section 2.6.2; Figure 5).

### 3.4 3D Trajectory Estimate

Using ballistic arrival travel times from 30 stations and the radar-derived terminal point, a best-fitting trajectory was determined with an average velocity of 25.27 km/s, azimuth of 300.36°, entry angle of 18.85° from horizontal, and terminal altitude of 23 km (Figure 4). The geographic entry point (onset of the luminous path) was located at 60.2229° N, 151.4665° W at an altitude of 79.3 km. The best-fitting celerity was 311 m/s, consistent with celerities used in the best-fitting solution for other bolide infrasound studies (Che et al., 2016; Walker et al., 2010). Under the constant-celerity (311 m/s) assumption, the largest deviation between observed and predicted travel times was <7 s among all stations, with a mean residual of <1.25 s (Figure 4a), indicating a robust solution despite the simplified atmospheric model.

The ballistic N-wave recorded at the KENI array yielded an independently derived point along the trajectory at 69 km altitude, located at 60.4016° N, 151.9574° W (Figure 5). This point was only 2.52 km from the corresponding location on the travel-time-based trajectory, providing strong independent validation of the inversion result (Figure 6a).

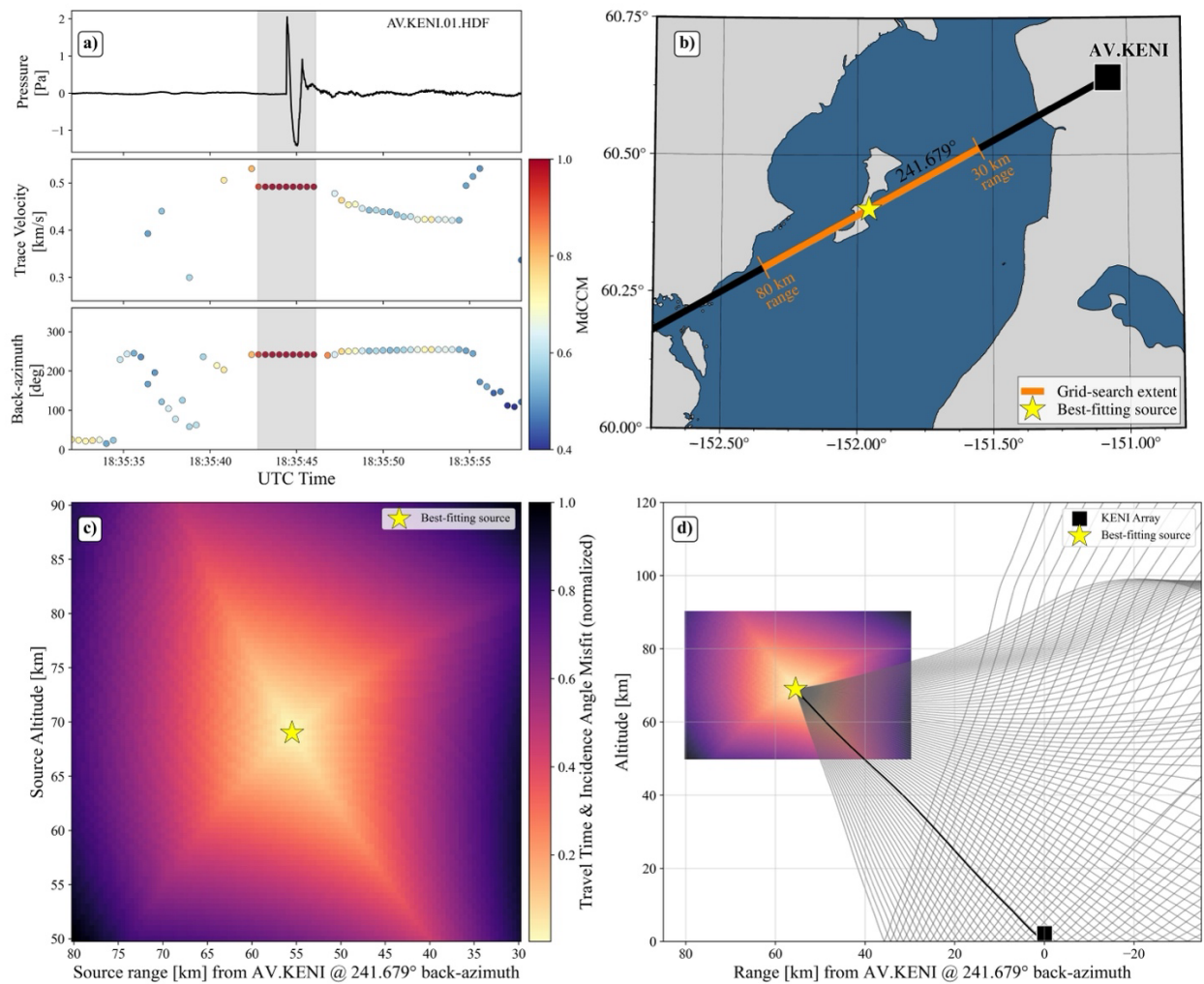

**Figure 5:** Estimates of a point along trajectory using a single infrasound array. a) Array processing results at the KENI infrasound array. Top panel contains the ballistic N-wave, middle panel is the trace (apparent) velocity, and the bottom panel is the back-azimuth estimate. Values are colored by the median cross-correlation maxima (MdCCM). b) Map showing the horizontal geographic extent of the grid search parameter space used to locate a point along the cylindrical line source. c) Heatmap of the misfit grid showing the best-fitting source range and altitude. d) InfraGA raypaths shot from the best-fitting source location towards the KENI array, where the bold line is the source-receiver ray. The misfit grid is also shown for spatial context.

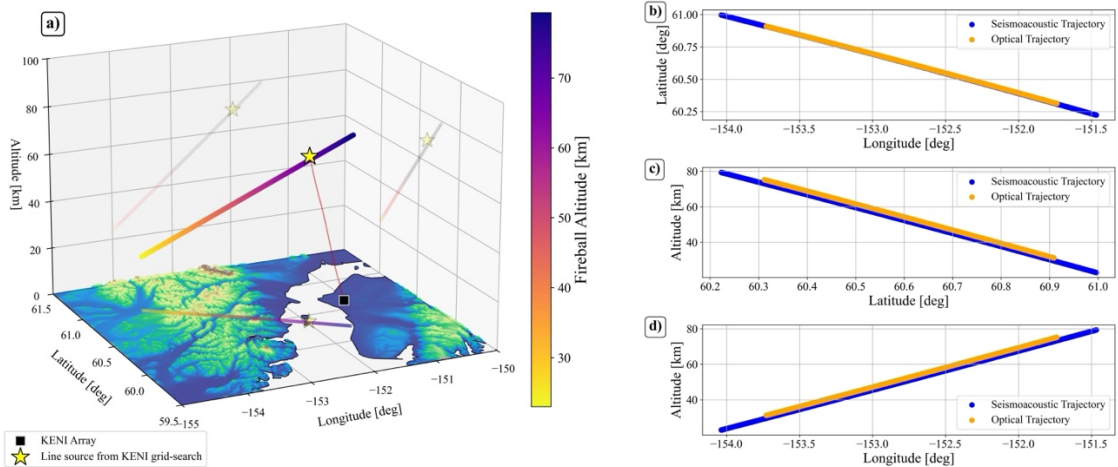

**Figure 6:** a) 3D view of the seismoacoustic-derived fireball trajectory, where the color represents fireball altitude, and the yellow star is the best-fitting location along trajectory using the KENI array ballistic detection (see Figure 5). An eigenray is shown connecting the yellow star and KENI. b), c), and d) Comparison of the optical trajectory solution (orange) and the seismoacoustic-derived trajectory (blue).

### 3.5 Energy Estimates

The averaged signal period across all stations was 1.50 ± 0.80 s. Period measurements at individual stations and arrays can be found in Tables S1 and S3 (SI). For ballistic arrivals only, dominant signal period was 1.11 ± 0.32 s. For arrivals attributed to fragmentation events, dominant signal period was 2.17 ± 0.97 s. These two signal classes reflect fundamentally different sampling of the fireball trail: ballistic arrivals capture the continuous line-source shock along the hypersonic path, whereas fragmentation arrivals record discrete energy releases at localized breakup points. As a result, they preserve distinct modes of energy deposition.

Applying the bolide-specific period–yield relationship using the averaged period across all stations (1.50 s) for both ballistic and fragmentation signals yielded an energy estimate of $3.825 \cdot 10^{-2}$ kt (38.25 tons) TNT equivalent. This approach is consistent with prior bolide infrasound yield studies that employ station-averaged dominant periods (e.g., Ens et al., 2012; Gi & Brown, 2017; Silber et al, 2025). If we include the uncertainty in the averaged period, then the energy ranges from 2.25 to 186.77 t TNT equivalent. However, we note that this range should be interpreted with caution, as the uncertainty in signal period is quite large because of a significant span in measured values due to fundamentally different energy deposition modes. Using the single averaged period energy estimate (38.25 t TNT) and the optical-derived entry velocity of 23.37 km/s (see Section 3.6), the object's infrasonically derived mass (Silber et al., 2015) was estimated at 586 kg. Assuming a chondritic bulk density of 3.36 ± 0.16 g/cm$^3$ this corresponds to a diameter range of ~0.68-0.70 m.

**3.6 Optical Results**

The independent optically derived trajectory is shown with the orange line in Figure 6b-d. The relative location of the two optical video observations are plotted in Figure S5 (SI). The optical trajectory analysis of the fireball (see Table S4 in SI) shows that it entered the atmosphere at a shallow 18.99° angle and a direction of flight azimuth of 302.3°. Trajectory fit errors are shown in Figure S6 (SI). At the last observed point, the fireball was at a height of 31.2 km and moving at ~20 km/s, decelerating slightly from its initial velocity of 23.37 km/s. At this last point, the fireball reached a dynamic pressure of ~6 MPa, indicating a rocky composition. The orbit parameters are listed in Table S5 (SI). The meteoroid was on an Apollo-type heliocentric orbit ($a$ = 1.65 AU, $e$ = 0.66) with an inclination of 10.72º, its perihelion at 0.57 AU and aphelion at 2.73 AU, placing its origin within the main asteroid belt.

Based on the measured orbital elements, an informed guess can be made about what type of meteorite may have fallen based on the orbital element distribution of previously documented meteorite falls and the overall fall statistics of observed meteorite falls of different types. Based on the still small sample of 75 cases discussed in Jenniskens & Devillepoix (2025), we calculated that the most likely type to have fallen here is an L-type ordinary chondrite (~44%), followed by an H chondrite from the Central Main Belt source (~30%), an HED achondrite (~13%), or an LL chondrite (~11%), respectively. It is unlikely that the fall was a CM carbonaceous chondrite or an H-type ordinary chondrite from either the Koronis or Inner Main Belt sources.

We compared our optically derived trajectory to the seismoacoustic solution (Figure 6b-d). At the beginning of the optical path (75.2 km altitude) the two trajectories are separated by 2.3 km, which gradually decreases to a minimum deviation of 0.57 km at 49.5 km altitude, and finally increases to a 3D distance of 1.7 km at the end of the optical trajectory (31.2 km altitude). The mean deviation between the two solutions along the entire path was 1.97 km.

**4. Discussion**

**4.1 Event Uniqueness and Multi-Sensor Data Integration**

The 24 April 2025 Southcentral Alaska fireball produced a complex seismoacoustic shock footprint, recorded by a densely instrumented geophysical network. While large bolides have been observed with dense seismic coverage before, such as the 2022 Central Italy bolide (Olivieri et al., 2023) and the 2008 Oregon bolide (Walker et al., 2010), it is rare for such an event to occur within range of a comparably dense infrasound network (Figure 1), especially one supplemented by Doppler weather radar signatures of falling meteorites (Figure 3). As opposed to a purely seismic network, the extensive acoustic recordings of the Alaska fireball allowed for improved source energy estimation because of better sampling of the path. Shallow atmospheric entries such as this one are statistically less common than steeper impacts (e.g. Ceplecha et al., 1998) but are particularly valuable scientifically, as they remain luminous and acoustically active for longer durations (Brown et al., 2013; Popova et al. 2013). Such trajectories allow extended observation of the coupled ablation, fragmentation, and shock-generation processes that are often compressed into much shorter spatio-temporal scales for steeper entries (e.g. Brown et al., 2003; Silber 2025; Silber et al., 2024b).

Unlike typical bolide studies, which rely on optical data to define the trajectory and use seismoacoustic observations as secondary constraints (McFadden et al., 2021; Pilger et al., 2020; Silber, 2024b) we inverted this paradigm. The initial trajectory was refined using seismoacoustic arrivals and infrasound array back-azimuth estimates, which lead to a targeted search in Doppler radar. After radar identification of the meteorite fall, this location (combined with timing information from a single optical observation) then seeded the final seismoacoustic-trajectory solution. Figure 7 aims to clarify this unique workflow and provides a chronology for the analysis performed in this study.

SAND2026-18925O

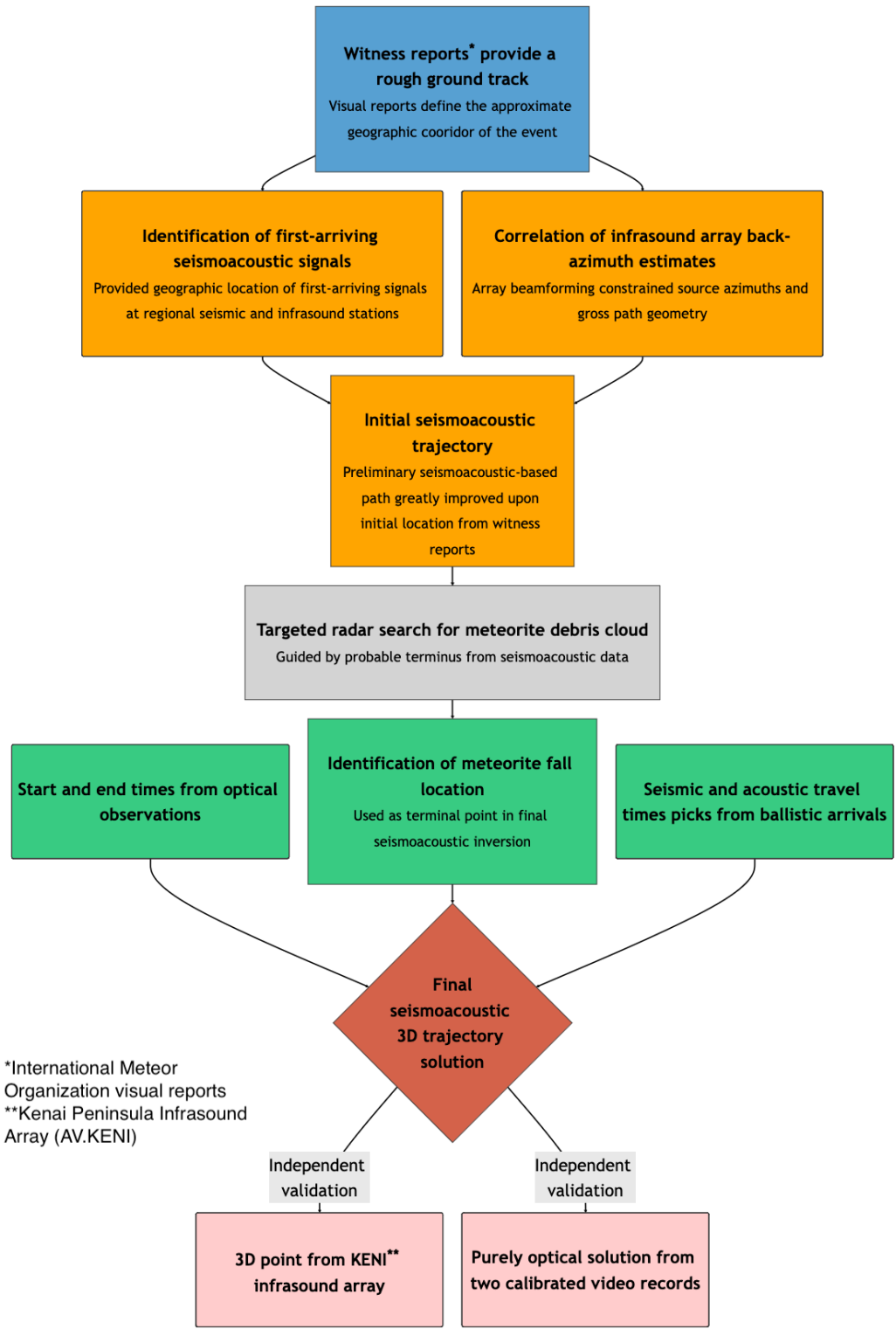

**Figure 7:** Chronological diagram of the workflow to derive the final seismoacoustic-based trajectory solution in this study. The chronology begins at the top and ends at the bottom.

Our framework is, to our knowledge, the first documented and validated demonstration of using seismoacoustic information to directly guide searches for a meteorite fall. Given that this event was undetected by space-based sensors, we can envision the use of seismoacoustic observations to identify future meteorite falls that may otherwise have gone unnoticed. By analyzing initial arrival times and tracking the propagation of acoustic signals across the sensor network, it is theoretically possible to gather sufficient timing and trajectory information to initiate a radar search, independent of optical observations.

Radar observations of falling meteorites in this study parallels recovery campaigns such as Sutter's Mill (Jenniskens et al., 2012) and Creston (Jenniskens et al., 2019), but with the main difference that recovery has not yet been attempted here due to the remote location. The fall is located in the Stony River valley, a glacial outflow channel with thick vegetation and abundant stream channels (Figure 3). Given the geographic setting, travel to and subsequent search efforts at this location would likely prove very difficult. However, if ever attempted, recovery chances would likely be best on westernmost side of the strewn field, where there is a topographic high and less vegetation and water. This western location also coincides with the largest meteorites deposited.

**4.2 Ballistic Arrivals and Fireball Trajectory**

Ballistic arrivals were recorded at 30 stations (Figure 4a), an unusually high number even compared to energetic events such as Chelyabinsk (Brown et al., 2013; Le Pichon et al., 2013; Popova et al., 2013), or the 2018 Bering Sea bolide (Arrowsmith et al., 2021). Ballistic arrivals are particularly valuable as they directly sample the line-source shock geometry of the fireball path, unlike fragmentation arrivals, which sample discrete energy releases (Edwards, 2009; ReVelle, 1976; Silber & Brown, 2014). In the present study, although the source was a stochastic natural event, the exceptional density of ballistic detections enabled a detailed sampling of the path similar to planned campaigns like the OSIRIS-REx sample return capsule reentry (Silber & Bowman, 2025).

During its traverse, the fireball passed nearly overhead Redoubt Volcano, which is heavily instrumented with seismic and infrasound sensors. As previously mentioned, a wide variability of infrasound amplitudes (2.4-14.3 Pa) over a particularly small distance (15.5 km) was observed. This large amplitude variation over short ranges may reflect fine-scale atmospheric structure causing azimuthally dependent propagation effects, a phenomenon that could be explored further using this dataset as a natural analog to calibration campaigns. We also speculate that should fragmentation occur at a location coincident with the ballistic (cylindrical line) source that is a recorded by a particular station, then one acoustic station may be sampling the amplitude of both fragmentation and hypersonic shock front. In this scenario, another nearby station could sample only the shock source, causing the significant amplitude variation.

The seismoacoustic-based 3D trajectory reconstruction of the Alaska fireball (Figures 4,6) integrated multi-station ballistic travel times, optical observations for timing, and a radar-derived terminal point (identified using initial seismoacoustic observations and infrasound arrays). For validation and comparison, the ballistic arrival at the KENI array enabled the location of a single point in 3D space (Figure 5). The ability to do this with only a single infrasound array and optical observation (for timing) is significant. While not presented in this study, one could envision the use of this point in space as a seed for the seismoacoustic inversion rather than the radar location. This would lead to a trajectory solution entirely independent of radar data, only requiring seismoacoustic signals and timing information from a single optical observation. While we do not explore this solution in this study, we do highlight this as a potential to reconstruct fireball trajectory with more limited information.

Optical observations produced an independent trajectory solution, providing an additional validation dataset to our seismoacoustic-based trajectory. The reconstructed seismoacoustic trajectory exhibited excellent spatial agreement with the optical solution, with a mean lateral deviation of only 1.97 km between the two paths (Figure 6b-d). Furthermore, velocity and azimuth estimates derived from the two approaches differ by ~2 km/s, and ~2°, respectively, well within the range of expected measurement and modeling uncertainties for independently constrained datasets. The entry angle estimates were near-identical, differing by only 0.2°, where the seismoacoustic-based trajectory was marginally less (18.85°) than the optical (18.99°). While the entry angles are similar, the seismoacoustic trajectory is consistently at a slightly lower altitude than the optical solution (Figure 6b-d). We posit that this is due to the constant celerity assumption, where the inversion is balancing the travel times between closer stations, which likely need faster celerities, and further stations which have slower celerities due to different propagation schemes. Nonetheless, this high level of agreement across two fundamentally different sensing modalities demonstrates the robustness and reliability of seismoacoustic approaches for reconstructing bolide flight paths with high spatial and kinematic accuracy. A future test for our seismoacoustic approach could include reconstruction of controlled reentries, such as OSIRIS-REx (Silber et al., 2024a), which have exceptionally well-constrained trajectories and would service as a useful near-ground-truth dataset for comparison.

### 4.3 Fragmentation Episodes and Multi-Modal Correlation

Video analysis revealed three distinct luminous fragmentation episodes toward the end of the trajectory. Infrasound array processing identified multiple acoustic sources along the path, with one well-constrained location at ~33 km altitude that coincided closely in both time and space with the second luminous episode (from optical). This agreement demonstrates the value of combining optical and acoustic observations for resolving fragmentation dynamics.

The correspondence between luminous and acoustic fragmentation is consistent with previous bolide studies that documented how catastrophic breakups are expressed in both optical and acoustic datasets (Hedlin et al., 2018). Events such as Peekskill (Beech et al., 1995), Morávka (Borovička & Kalenda, 2003), and Košice (Borovička et al., 2013) showed that major luminous flares correspond to fragmentation episodes that release significant acoustic energy. Laboratory-calibrated and observational studies of atmospheric fragmentation (Ceplecha et al., 1993; Henych et al., 2023) further confirm that fragmentation cascades produce multiple discrete episodes rather than a single terminal disruption.

Our result is also consistent with theoretical expectations that fragmentation episodes occurring under dynamic pressures of ~0.5–5 MPa often dominate the acoustic energy budget (Pilger et al., 2020; Silber et al., 2025). The altitude at which these pressures occur depends on entry velocity and atmospheric density, but for meter-class chondritic bodies such breakups typically occur between 25 and 40 km (Borovička et al., 2020). The ~33 km fragmentation altitude inferred here fits squarely within this range, indicating that the Alaska fireball fragmented under conditions consistent with other well-studied bolide falls.

**4.4 Acoustic-Seismic Coupling and Precursory Waves**

A notable feature of this dataset is the possible presence of precursory seismic signals observed at several stations (Figure S3). These arrivals precede the direct acoustic wave and are most plausibly explained as surface waves activated by infrasound–seismic coupling at locations offset from the recording sites. In this mechanism, the apparent (trace) velocity of the acoustic wavefront matches the phase velocity of leaky-mode or Rayleigh waves in the solid Earth, producing efficient energy transfer, narrowband spectral content, and waveforms resembling monochromatic surface waves (Xu & Lognonné, 2024). Because the apparent velocity that induces these waves exceeds the acoustic velocity, they can arrive at separate stations before the acoustic wave. A characteristic of these waves is their elliptical particle motion (Press & Ewing, 1951) which we illustrate in Figure S3e, supporting our interpretation. Elevated sources such as bolides are favorable for this process because their steeply downgoing or oblique wavefronts have higher apparent velocities than horizontally propagating acoustic waves, increasing the likelihood of mode conversion into seismic surface waves.

Observations of such precursory seismic waves from atmospheric entries remain rare. The 1908 Tunguska explosion produced Rayleigh waves recorded up to 5293 km from the source (Ben-Menahem, 1975). Similar precursory signals have been reported by Anglin and Haddon (1987) from a meteoroid above Yellowknife, Canada, and Brown et al., 2003 from the Morávka fireball. Tauzin et al. (2013) demonstrated unambiguous seismoacoustic coupling associated with the breakup of the Chelyabinsk meteor. Comparable phenomena have also been documented during spacecraft re-entries, including the Columbia space shuttle (Kanamori et al., 1991; Sorrells et al., 2002). Possible precursor waves were reported in the context of Hayabusa re-entry (Ishihara et al., 2012).

Within this context, the Alaska fireball provides one of the few modern examples of precursory surface-wave activation linked to bolide infrasound. These new observations expand the empirical basis for understanding seismoacoustic coupling processes and highlight the diagnostic value of dense regional networks for capturing rare phenomena at the interface between atmospheric and solid-Earth wavefields.

### 4.5 Period Variability and Energy Deposition

Substantial variability in dominant periods of bolide-generated infrasound has been documented in numerous events, with explanations ranging from atmospheric propagation effects to differences in which portion of the trajectory a given station samples (Ens et al., 2012; Silber et al., 2015; Pilger et al., 2020). Silber et al. (2011), in their analysis of the 2009 Indonesian bolide, applied blast-radius modeling to show that short-period signals are associated with cylindrical line-source shocks generated in the upper trajectory, whereas longer-period signals originate from lower altitudes where the blast cavity radius $R_0$ is larger, particularly during terminal fragmentation and airburst. This interpretation is consistent with weak-shock theory (ReVelle, 1974, 1976), where the fundamental period is linked to $R_0$.

The Alaska fireball provides a uniquely well-constrained case to test these hypotheses. The shallow entry angle produced an extended atmospheric path, and the dense regional seismoacoustic coverage enabled separation of ballistic arrivals from fragmentation arrivals. Our observation that acoustic signal periods from ballistic arrivals (1.11±0.32 s) and fragmentation arrivals (2.17±0.97 s) are systematically different has important implications for bolide energy estimation. In principle, period-yield relationships (Ens et al., 2012; Gi & Brown, 2017; ReVelle, 1997) assume all stations sample the same source regime. Silber et al. (2025) postulated that period-yield relations might not be one-size fits all and that bolide parameters, including the mode of energy deposition, must be considered. Indeed, our data show stations detected either ballistic shocks or fragmentation episodes depending on geometry. This could introduce systematic uncertainty into single-station or sparse- network yield estimates, as they may only sample one of the underlying processes generating acoustic signals. For example, because fragmentation signals are generally lower frequency, they may be able to propagate longer distances and could bias energy estimation without sufficient sampling of the ballistic source regime.

Given this context, application of period–yield relationships is most robust when using averaged periods across all contributing stations (Ens et al., 2012; Gi & Brown, 2017; Silber et al., 2025). Averaging provides a statistically stable metric that reduces scatter caused by variable sampling of different trail segments or shock regimes, while still capturing the underlying source physics.

### 4.6 Implications for Planetary Defense and Hazard Monitoring

This event was not detected by USG sensors or the Geostationary Lightning Mapper (GLM) aboard Geostationary Operational Environmental Satellites (GOES), illustrating a gap in the current global monitoring architecture for bolide events, with implications for planetary defense and near-Earth object (NEO) hazard assessment. Space-based optical and radiometric systems provide valuable coverage for high-yield events but often face limitations in detecting smaller or lower-luminosity bolides due to detection thresholds, orbital geometries, and atmospheric constraints, particularly in high-latitude regions or during daytime. The lack of space-based detection for this fireball is likely due to a combination of factors. For the GLM, the event occurred at a high latitude (60.9° N), which is at the extreme edge of the GOES-West instrument's field of view where sensitivity is significantly diminished. For USG, the combination of high-latitude geometry and daytime illumination, which increases background solar radiance, may have hindered detection of this relatively low-luminosity event. This event also reveals a broader observational gap at high latitudes, where space-based optical sensors operate at extreme viewing angles and ground-based networks are comparatively sparse. Much of the Arctic and sub-Arctic region remains poorly instrumented for fireball monitoring, creating challenges for timely detection and characterization of small to medium-sized impactors.

In contrast to space-based platforms, this event was well-recorded by a dense network of ground-based seismoacoustic sensors, and a few casual video records. These ground-based capabilities complement space-based systems by filling observational gaps, improving event characterization, and expanding the detection envelope for smaller energies. The high density of directional observations in this case was aided by the fireball's shallow entry angle, which extended the atmospheric path length and increased the range of favorable viewing geometries for stations along multiple azimuths. In planetary defense terms, such entry geometries can provide extended opportunities for multi-azimuth detection, improving both the spatial sampling of ballistic shock fronts and the precision of trajectory reconstruction.

While infrasound monitoring offers global coverage and the ability to detect events in remote or obscured regions, its delayed signal arrival limits its use for early warning. Instead, we show that seismoacoustic methods paired with additional datasets could excel in rapid post-event characterization, enabling determination of trajectory, energy yield, fragmentation behavior, and aid in the location of meteorite falls. For planetary defense applications, such rapid reconstruction is critical for situational awareness and hazard assessment in the minutes to hours following an impact.

The estimated yield of ~38.25 t TNT equivalent is comparable to several recovered meteorite falls (e.g., Sutter's Mill; Jenniskens et al., 2012) but occurred at an unusually high entry velocity (~23.37 km/s) for meter-class impactors (Ceplecha et al., 1998; Drolshagen et al., 2020). High-velocity entries produce stronger shock waves and can increase detectability at greater ranges, which could contribute to the large number of seismoacoustic detections we found for this event.

The methodology demonstrated here can also apply to space debris and artificial object reentries. Sufficiently large spacecraft components and sample return capsules can generate acoustic signatures similar to bolides (Clemente et al., 2025; Ishihara et al., 2012; Yamamoto et al., 2011; Bowman et al., 2025). For example, previous studies of planned reentries (Nishikawa et al., 2022) and several controlled and uncontrolled spacecraft returns (Garcés et al., 2004; de Groot-Hedlin et al., 2008; McKenna & Herrin, 2006) have shown that seismoacoustic networks are capable of constraining fragmentation altitude, trajectory, and breakup dynamics when optical coverage is limited or predictions are uncertain.

By demonstrating that high-accuracy, multi-sensor trajectory reconstruction can be achieved opportunistically for a stochastic natural bolide, this study bridges the methodological gap between planetary defense and space traffic management. The combination of dense infrasound and seismic coverage, along with radar observations guided by the infrasound solution, provides an independent terminal constraint. This robust trajectory inversion model enables a scalable model for rapid post-event characterization across both domains, ensuring that even when an entry cannot be predicted, it can be understood quickly and accurately.

## 5. Conclusions

The 24 April 2025 Southcentral Alaska fireball produced one of the most densely recorded seismoacoustic datasets ever obtained for a natural bolide. Thirty ballistic arrivals were captured across a regional network of seismic and infrasound sensors, complemented by Doppler weather radar detections of falling meteorites and independent optical video observations. This convergence of multi-sensor data provided a rare opportunity to reconstruct a stochastic atmospheric entry with high fidelity and to evaluate fragmentation behavior.

By integrating optical timing, infrasound, seismic, and radar datasets, we obtained a trajectory solution with a mean travel-time residual under 1.25 s, yielding an entry angle of 19°. After the seismoacoustic trajectory solution was obtained, it was cross-validated by the KENI array's estimate of a single point, and a purely optical trajectory solution. The optical trajectory solution showed an excellent agreement with the seismoacoustic-based solution, deviating by an average of <2 km in 3D space. The optically-derived velocity was determined to be 23.37 km/s, within 2 km/s of the seismoacoustic estimate, further reinforcing the robustness of the reconstruction. Using orbital parameters from the optical trajectory solution, we estimate the object's composition as most likely a L-type ordinary chondrite. The event released an estimated ~38 t TNT equivalent of kinetic energy. Assuming a chondritic composition, the pre-entry body a diameter of ~0.7 m. Radar signatures of falling debris indicate that significant (up to 1 kg) meteorites were deposited. While no on-site investigation has been made yet due to the remote location, recovery may be possible in the future. Optical video documented three luminous fragmentation episodes, one of which coincided spatially and temporally with a primary seismoacoustic fragmentation source, directly linking acoustic and optical breakup signatures.

This event demonstrates the operational value of dense ground-based networks as a complementary capability to space-based assets within the planetary defense framework. This event occurred in a sub-Arctic region, an area with limited space-based observational coverage, demonstrating the critical role of dense ground-based seismoacoustic networks in characterizing high-latitude atmospheric entries. Although infrasound cannot provide early warning due to acoustic propagation delays, its integration with seismic and optical/radar measurements could enable rapid, high-precision post-event characterization of unpredicted entries. The Alaska fireball shows that, when such complementary systems are available, a natural bolide could be reconstructed with calibration-grade accuracy, yielding trajectory, energy, and fragmentation parameters within hours. This approach advances both scientific analysis and the operational capacity to characterize future atmospheric entries, whether natural impactors or artificial reentries.

Scamfer, L. T., Silber, E. A., Fries, M. D., Vida, D., Šegon, D., Jenniskens, P., Nishikawa, Y., Sawal, V., & Rector, T.A. (2026). Multi-Sensor Trajectory Reconstruction of the 24 April 2025 Alaska Fireball and Implications for Planetary Defense. *Journal of Geophysical Research: Planets*. 131 , e2025JE009440. https://doi.org/10.1029/2025JE009440

**Funding:** This work was supported by the Nuclear Arms Control Technology (NACT) program at the Defense Threat Reduction Agency (DTRA). DV was funded for this work in part through NASA cooperative agreement 80NSSC24M0060.

**Conflict of interest:** The authors declare no conflict of interest.

SAND2026-18925OChe, I. Y., Kim, G., & Lee, H. I. (2016). Seismic and infrasonic analysis of the 9 March 2014 fireball in South Korea. *Geosciences Journal, 20*(2), 209–220. https://doi.org/10.1007/s12303-015-0034-1

Christie, D. R., & Campus, P. (2009). The IMS infrasound network: Design and establishment of infrasound stations. In A. Le Pichon, E. Blanc, & A. Hauchecorne (Eds.), *Infrasound Monitoring for Atmospheric Studies* (pp. 29–75). Springer Netherlands.

Clemente, I., Sansom, E. K., Devillepoix, H. A., Kawamura, T., Fernando, B. A., Garcia, R. F., & Collet, O. (2025). Exploring seismic signal detection and source identification of atmospheric entries: The Hayabusa2 sample return capsule as a benchmark. *Seismological Research Letters, 96*(5), 2780-2795. https://doi.org/10.1785/0220250032

Cohen, A. N., Lubin, P., Robertson, D., Boslough, M., Egan, S., Stickle, A. M., & Patel, D. (2024). Asteroid disruption and deflection simulations for multi-modal planetary defense. *Acta Astronautica, 225*, 960–967.

Consolmagno, G. J., Britt, D. T., & Macke, R. J. (2008). The significance of meteorite density and porosity. *Geochemistry, 68*(1), 1–29. https://doi.org/10.1016/j.chemer.2008.01.003

Devillepoix, H. A. R., Bland, P. A., Sansom, E. K., Towner, M. C., Cupák, M., Howie, R. M., & Cox, M. A. (2019). Observation of metre-scale impactors by the Desert Fireball Network. *Monthly Notices of the Royal Astronomical Society, 483*(4), 5166–5178. https://doi.org/10.1093/mnras/sty3442

Devillepoix, H. A. R., Cupák, M., Bland, P. A., Sansom, E. K., Towner, M. C., Howie, R. M., & Tate, J. R. (2020). A global fireball observatory. *Planetary and Space Science, 191*, 105036. https://doi.org/10.1016/j.pss.2020.105036

Drob, D. P., Picone, J. M., & Garcés, M. (2003). Global morphology of infrasound propagation. *Journal of Geophysical Research: Atmospheres, 108*(D21). https://doi.org/10.1029/2002JD003307

Drolshagen, E., Ott, T., Koschny, D., Drolshagen, G., Schmidt, A. K., & Poppe, B. (2020). Velocity distribution of larger meteoroids and small asteroids impacting Earth. *Planetary and Space Science, 184*, 104869. https://doi.org/10.1016/j.pss.2020.104869

DuMond, J. W. M., Cohen, E. R., Panofsky, W. K. H., & Deeds, E. (1946). A determination of the wave forms and laws of propagation and dissipation of ballistic shock waves. *The Journal of the Acoustical Society of America, 18*(1), 97–118. https://doi.org/10.1121/1.1916347

Edwards, W. N. (2009). Meteor generated infrasound: Theory and observation. In A. Le Pichon, E. Blanc, & A. Hauchecorne (Eds.), *Infrasound Monitoring for Atmospheric Studies* (pp. 361–414). Springer Netherlands.

Edwards, W. N., Eaton, D. W., & Brown, P. G. (2008). Seismic observations of meteors: Coupling theory and observations. *Reviews of Geophysics, 46*(4). https://doi.org/10.1029/2007RG000253

Ehlert, S., & Erskine, R. B. (2020). Measuring fluxes of meteor showers with the NASA All-Sky Fireball Network. *Planetary and Space Science, 188*, 104938. https://doi.org/10.1016/j.pss.2020.104938

Ens, T. A., Brown, P. G., Edwards, W. N., & Silber, E. A. (2012). Infrasound production by bolides: A global statistical study. *Journal of Atmospheric and Solar-Terrestrial Physics, 80*, 208–229. https://doi.org/10.1016/j.jastp.2012.01.018
Scamfer, L. T., Silber, E. A., Fries, M. D., Vida, D., Šegon, D., Jenniskens, P., Nishikawa, Y., Sawal, V., & Rector, T.A. (2026). Multi-Sensor Trajectory Reconstruction of the 24 April 2025 Alaska Fireball and Implications for Planetary Defense. *Journal of Geophysical Research: Planets*. 131 , e2025JE009440. https://doi.org/10.1029/2025JE009440

*Journal of Geophysical Research: Planets*

Supporting Information for

# Multi-Sensor Trajectory Reconstruction of the 24 April 2025 Alaska Fireball and Implications for Planetary Defense

L. T. Scamfer[1,2*], E. A. Silber[1], M. D. Fries[3], D. Vida[4], D. Šegon[5], P. Jenniskens[6,7], Y. Nishikawa[8], V. Sawal[1], T.A. Rector[9]

[1]Sandia National Laboratories, Albuquerque, NM 87123, USA

[2]Geophysical Institute, University of Alaska Fairbanks, Fairbanks, AK 99775, USA

[3]Astromaterials Research and Exploration Science Division, NASA Johnson Space Center, Houston, TX 77058, USA

[4]Department of Physics and Astronomy, Western University, London, ON N6A 3K7, Canada

[5]Astronomical Society Istra Pula, Park Monte Zaro 2, Pula HR-52100, Croatia

[6]SETI Institute, Mountain View, CA 94043, USA

[7]NASA Ames Research Center, Moffett Field, CA 94035, USA

[8]Division of Math, Sciences, and Information Technology in Education, Osaka Kyoiku University, Osaka, Japan

[9]Department of Physics & Astronomy, University of Alaska Anchorage, Anchorage, AK 99508, USA

**Contents of this file**

    Text S1

    Figures S1 to S6

    Tables S1 to S5

    References

**Introduction**

This supporting information contains text, figures and tables. The text contains methods and results for an alternative trajectory inversion procedure detailed in Nishikawa et al., (2022). This alternative inversion scheme does not integrate radar data as described in the main text. Primarily, the figures support methods, discussion, and conclusions drawn in the text and give additional context to readers. Tables include information about seismoacoustic stations that detected the fireball, with various relevant signal measurements that are used in the manuscript. They also contain the optical trajectory solution and derived orbital parameters.

**Text S1: Alternate Trajectory Inversion**

As a validation of our primary inversion method in the main text Section 2.6.1, we tried a similar, but alternate scheme using the approach described in Nishikawa et al., (2022), applied to the observed arrival times of the ballistic Mach cone (ReVelle, 1976; Silber et al., 2015). As opposed to the initial method treating atmospheric celerity as a variable parameter, the celerity here was fixed at a mean value of 305 m/s, calculated from the atmospheric temperature profile using the ideal gas relation. The straight-line trajectory model was parameterized by five variables: the fireball's speed, arrival direction (azimuth), entry angle, and the ground-intersection point (latitude and longitude). A grid search was performed over the following ranges and steps: (1) speed (15–60 km/s in 2 km/s increments), (2) azimuth (290–320° in 1° steps), (3) entry angle (10-50° in 0.5° steps), and (4) ground-intersection point (±300 km in a geographic plane from the meteorite fall in 10km steps). This method is different from the initial method as the terminus is not fixed by the radar data of the meteorite fall. The objective of this analysis was to see how radar-anchoring improves the solution and explore any differences in results.

For each trial trajectory, the expected arrival times at all stations were computed from the closest-approach points along the path using Mach cone geometry (ReVelle, 1976) and the fixed mean sound speed. The best-fit parameters were determined by minimizing a weighted chi-square ($X^2$) misfit between observed and predicted arrival times, using only arrivals identified as ballistic.

The results of this approach produced a trajectory roughly consistent with the main text seismoacoustic reconstruction in terms of path geometry. However, compared to our main reconstruction (including the radar-derived anchor point) the minimum chi-square misfit results overall showed greater travel time-residuals (RMS residual ~4 s) and quite a high fireball velocity (49 km/s). While the minimum chi-square result shows a high velocity, the misfit surface of acceptable solutions ($X^2<1$) remains relatively flat in the 20-50 km/s range. Previous work using this inversion scheme has shown that trajectory azimuth and elevation are much better constrained than the absolute velocity, so we are not surprised by this difference between this and the optical solution (23.37 km/s). While these uncertainties exist with respect to the velocity constraint, the resulting entry angle (18.7°) and azimuth (300.9°) estimates remain quite similar to the optical solution (main text Section 3.6). Comparing this independent result with the inversion used in the main text highlights the utility of data assimilation and the radar-anchoring, which in this case, better constrained the velocity.

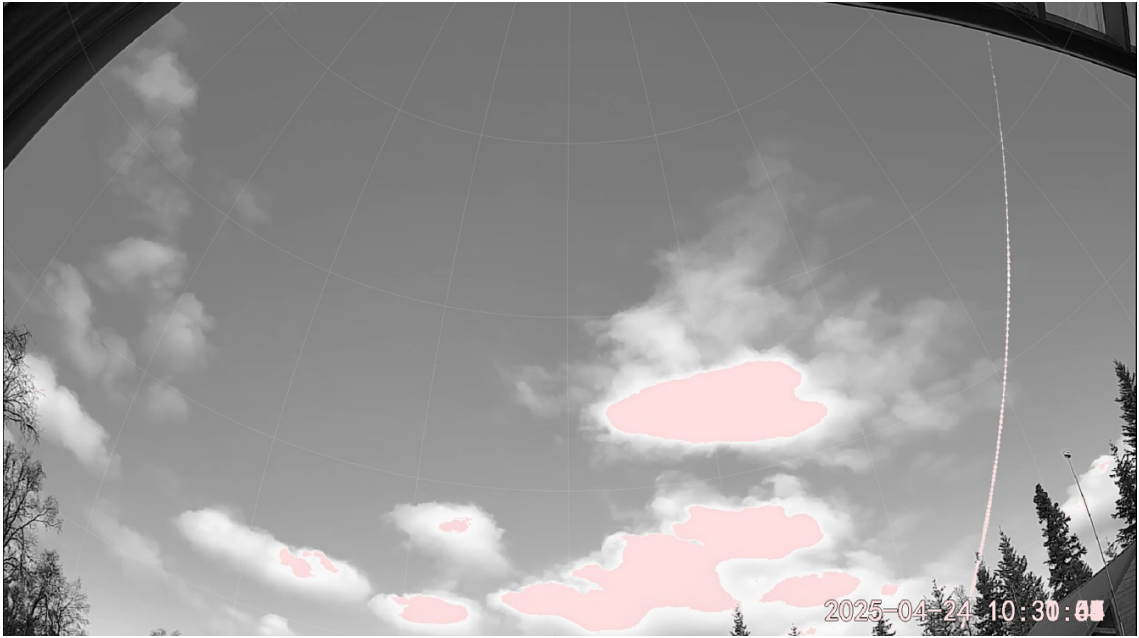

**Figure S1.** Calibrated optical image stack for Camera 1 near West Anchor Point, AK. This shows the fireball's path in the sky on the right side of the image.

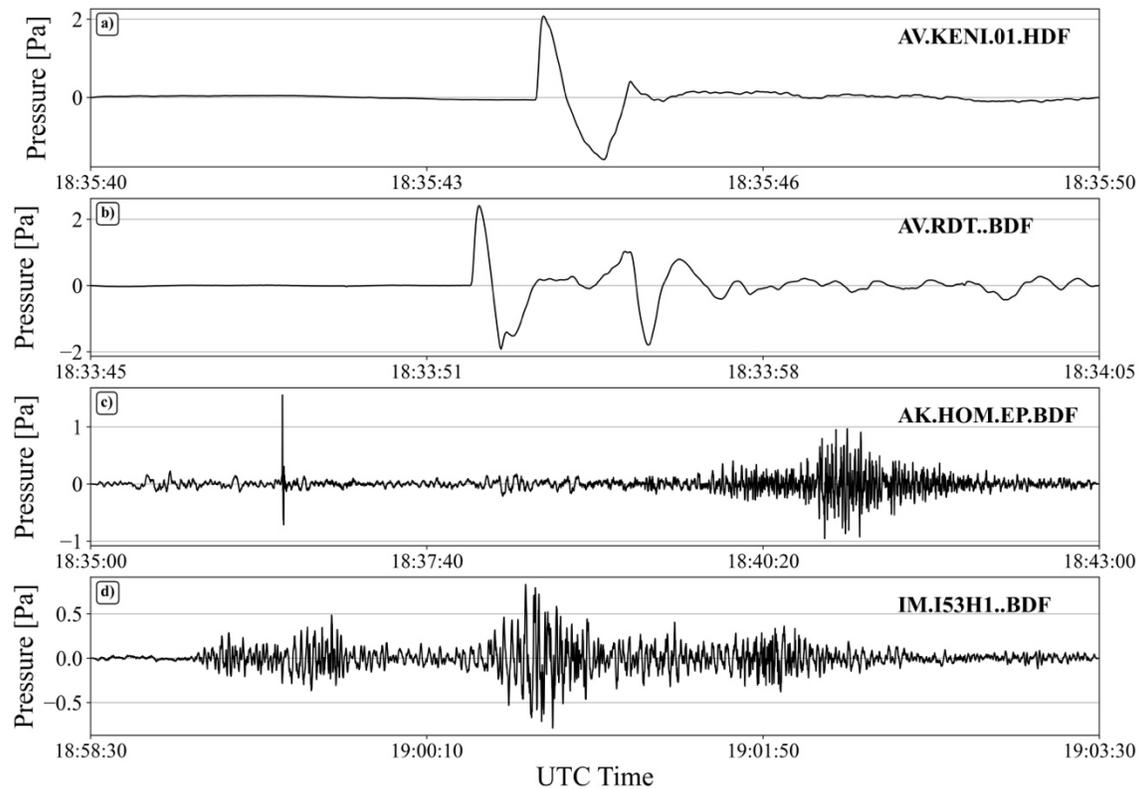

**Figure S2.** Examples of infrasound waveform diversity from fireball-related signals. Note that timescales are not equal between panels. Waveforms were bandpass filtered from 0.2-10Hz. a) N-wave recorded by an element of the KENI infrasound array b) Double N-wave recorded at infrasound station AV.RDT. c) Initial N-wave and subsequent emergent infrasound, likely from fragmentation episodes observed at AK.HOM. d) Infrasound from fragmentation episodes seen as packets at an element of IMS station I53US.

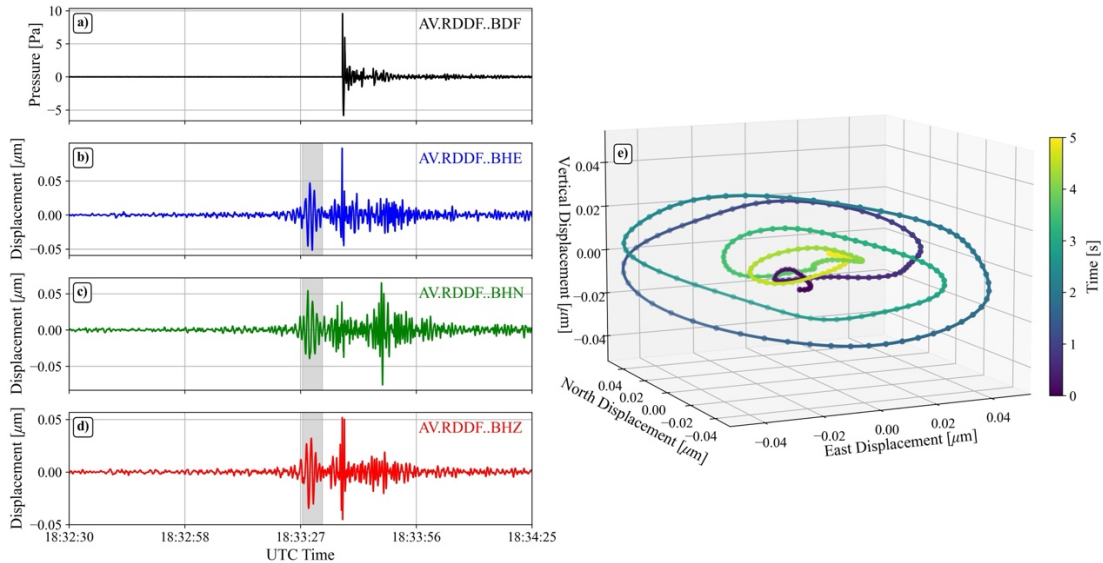

**Figure S3.** Infrasound (a) and seismic waveforms (b-d) at AV.RDDF, and particle motion (e) of a potential precursory seismic wave. Waveforms are filtered 1-10 Hz. The grey highlighted section is the time span used for particle motion analysis on the right. Seismic particle motion shown in panel e) has been interpolated for improved visual clarity; larger circles represent actual data points.

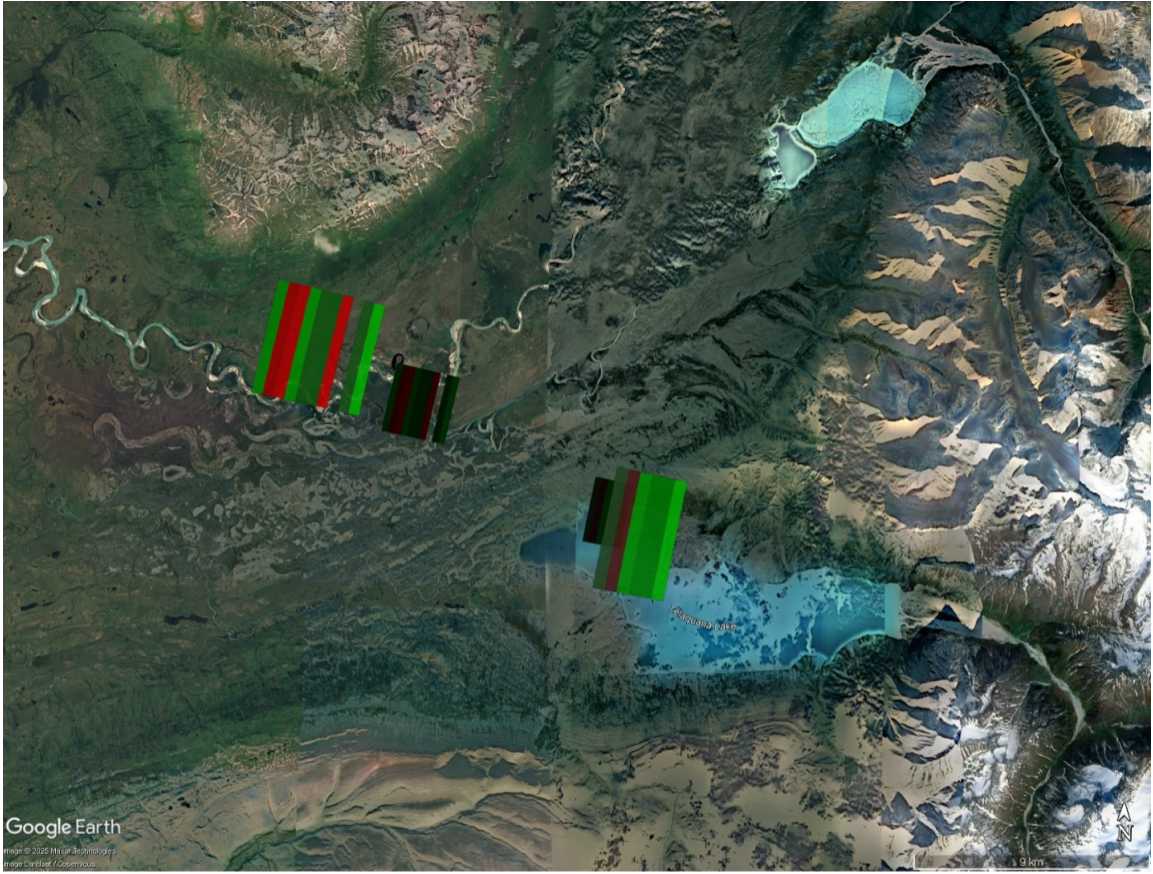

**Figure S4.** Doppler velocity data corresponding to the reflectivity values in Figure 3 in the main text, showing the "candy striping" signature typical of meteorite falls. Green indicates air motion towards the radar and red away from it. The strong values of intermixed direction shown here indicate short-range turbulence produced in the wake of fast-moving meteorites, probably in the 100s of grams and larger mass range. Modeling with the Jörmungandr dark flight model (Fries, 2023) placed the centroid of the strewn field at 60.9964° N, 154.0359° W. This centroid represents the observable mass-weighted debris distribution following dark flight. Because fragment descent depends on assumed fragment mass, starting altitude, and wind structure, dark-flight trajectories do not yield a unique luminous terminal coordinate but rather a distributed impact field. In this study, the radar-derived centroid was adopted as the terminal constraint for trajectory inversion (Section 3.4) because it is the only directly observable ground-impact indicator. Any attempt to specify a single deterministic terminal point would introduce additional model dependence beyond what is constrained by the radar observations. The centroid therefore provides a physically realistic anchor consistent with the inherent uncertainty in dark-flight modeling.

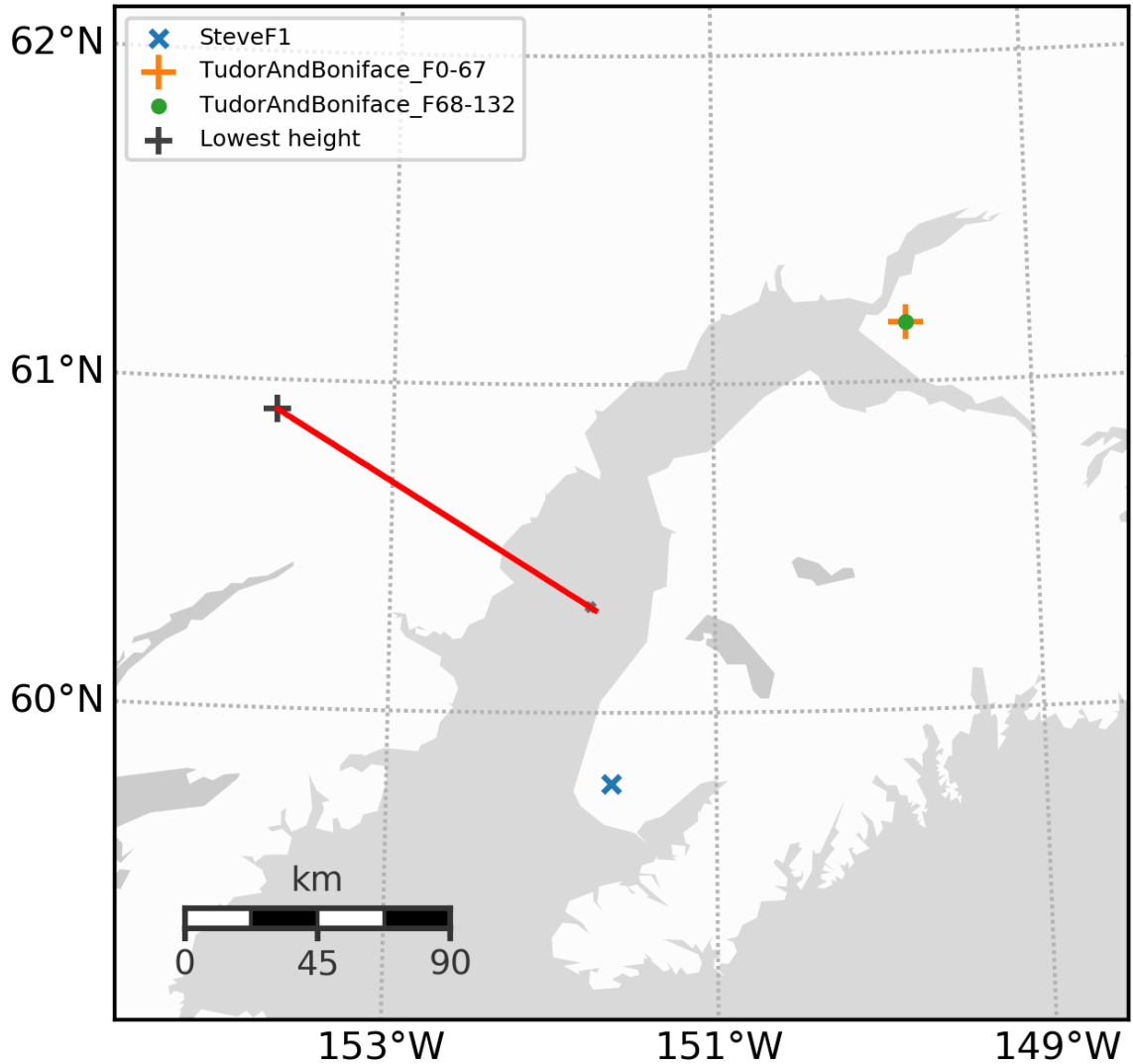

**Figure S5.** Location of the optically derived trajectory ground track relative to the Anchor Point camera (blue X) and dash camera video in Anchorage, AK (orange plus-sign). These two optical observations were used to derive an independent optical trajectory.

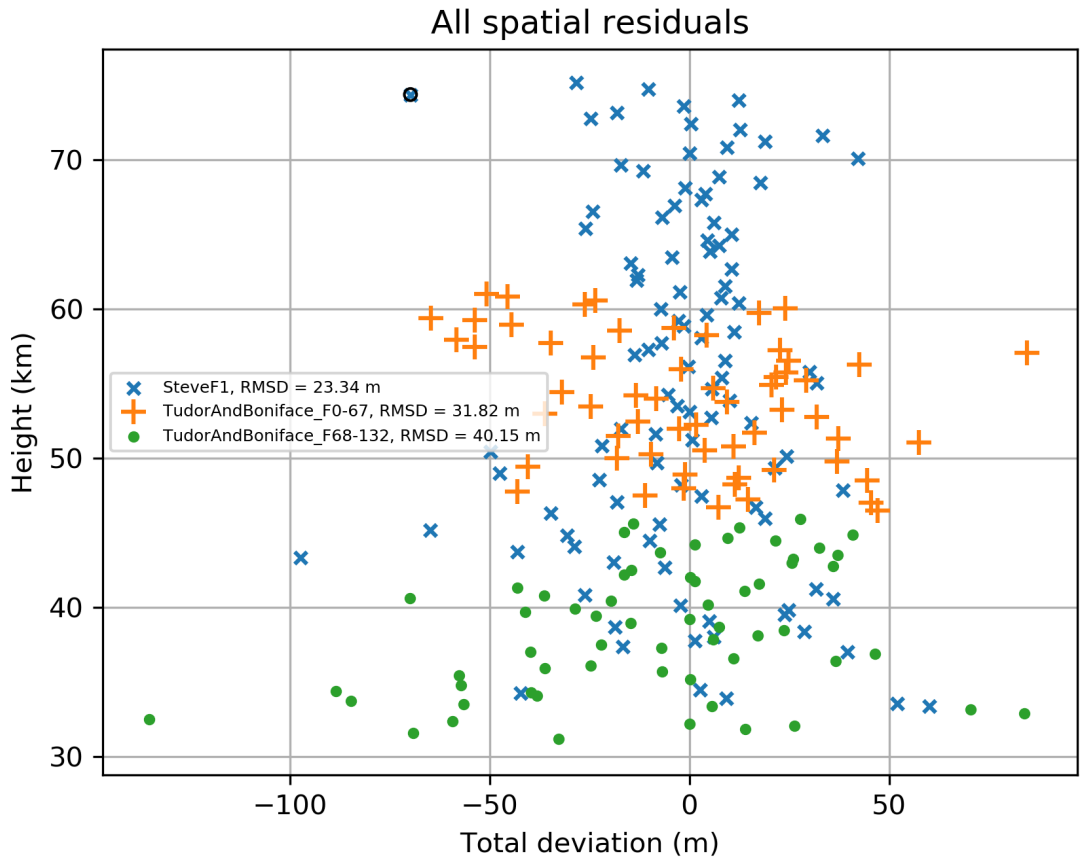

**Figure S6.** Trajectory fit errors for the optical trajectory versus fireball height. Due to a slight shift in the camera pointing, the dashcam video had to be broken up into two parts (at video frame 68). The formal trajectory fit errors were below 50 m, making the trajectory fit accurate.

**Table S1.** Overview of infrasound characteristics for stations that recorded fireball-related signals. FDSN is the International Federation of Digital Seismograph Networks. The infrasound signal period was measured at maximum amplitude using the 4 zero-crossings method (see Section 2.3 in the main text; Ens et al. (2012); ReVelle (1997); Silber (2024a)).

| FDSN Station | Station Latitude [deg] | Station Longitude [deg] | Station Altitude [m] | 3D distance to optical trajectory [km] | Signal onset time [UTC] | Ballistic arrival [Y/N] | Signal period [s] | Maximum amplitude [Pa] | Peak-to-peak amplitude [Pa] |
|---|---|---|---|---|---|---|---|---|---|
| AU22 | 59.370595 | -153.35506 | 105 | 150.93 | 18:39:02.45 | Y | 1.109 | 4.296 | 6.725 |
| AULG | 59.3773 | -153.5193 | 18 | 153.62 | 18:39:09.00 | Y | 1.119 | 3.724 | 6.691 |
| AUNO | 59.403992 | -153.451444 | 31 | 149.71 | 18:38:58.25 | Y | 1.118 | 4.421 | 7.425 |
| DORN | 61.067727 | -148.165709 | 22 | 225.35 | 18:47:14.00 | N | 1.932 | 0.181 | 0.302 |
| HOM | 59.6572 | -151.6515 | 221 | 104.69 | 18:36:31.25 | Y | 1.060 | 1.289 | 2.032 |
| O18K | 59.8542 | -155.208 | 544 | 146.34 | 18:39:13.00 | N | 3.778 | 0.173 | 0.405 |
| O19K | 60.1952 | -154.3201 | 96 | 91.38 | 18:35:58.75 | Y | 1.083 | 10.061 | 16.277 |
| P18K | 59.3922 | -155.2292 | 680 | 190.90 | 18:41:27.00 | N | 2.396 | 1.985 | 3.772 |
| P19K | 59.6524 | -153.2319 | 629 | 122.03 | 18:37:30.40 | Y | 0.652 | 3.606 | 7.880 |
| RC01 | 61.0889 | -149.739 | 390 | 158.08 | 18:39:25.80 | Y | 1.274 | 0.574 | 1.058 |
| RDDF | 60.5912 | -152.6883 | 1134 | 50.21 | 18:33:37.90 | Y | 0.573 | 14.339 | 20.575 |
| RDSO | 60.4536 | -152.7453 | 1557 | 53.92 | 18:33:50.10 | Y | 0.632 | 6.598 | 11.116 |
| RDT | 60.5726 | -152.4079 | 945 | 55.22 | 18:33:52.56 | Y | 1.520 | 2.430 | 4.313 |
| S15K | 56.3058 | -158.54 | 17 | 584.56 | 19:02:54.00 | N | 2.839 | 0.290 | 0.551 |
| SPCP | 61.265465 | -152.155495 | 1640 | 89.92 | 18:35:48.70 | Y | 1.244 | 8.518 | 15.639 |
| SPWE | 61.27346 | -152.559 | 1317 | 77.81 | 18:35:11.35 | Y | 1.449 | 11.745 | 21.097 |

**Table S2.** Overview of seismic characteristics for stations that recorded a fireball signal. Note that the table overflows to the next page.

| FDSN Station | Station Latitude [deg] | Station Longitude [deg] | Station Altitude [m] | 3D distance to optical trajectory [km] | Signal onset time [UTC] | Ballistic arrival [Y/N] |
|---|---|---|---|---|---|---|
| AU22 | 59.370595 | -153.35506 | 105 | 150.93 | 18:39:02.45 | Y |
| AUJA | 59.33956 | -153.42258 | 376 | 155.12 | 18:39:16.45 | Y |
| AUJK | 59.34593 | -153.41057 | 359 | 154.28 | 18:39:13.00 | Y |
| AUL | 59.38218 | -153.43765 | 327 | 151.39 | 18:39:04.00 | Y |
| AULG | 59.3773 | -153.5193 | 18 | 153.62 | 18:39:09.00 | Y |
| AUNO | 59.403992 | -153.451444 | 31 | 149.71 | 18:38:58.25 | Y |
| AUSB | 59.333 | -153.4284 | 230 | 155.91 | 18:39:18.75 | Y |
| AUW | 59.369413 | -153.47295 | 300 | 153.32 | 18:39:09.67 | Y |
| AUWS | 59.35851 | -153.46115 | 465 | 154.05 | 18:39:10.84 | Y |
| BRLK | 59.7511 | -150.9063 | 409 | 107.86 | 18:36:18.50 | Y |
| CCB | 64.6453 | -147.8053 | 233 | 514.93 | 18:58:01.00 | N |
| HDA | 64.4091 | -146.9482 | 437 | 519.31 | 18:58:15.00 | N |
| HOM | 59.6572 | -151.6515 | 221 | 104.69 | 18:36:31.25 | Y |
| L22K | 62.5258 | -150.7703 | 817 | 239.06 | 18:43:39.00 | N |
| MDM | 64.9602 | -148.2319 | 631 | 531.30 | 18:59:02.00 | N |
| O18K | 59.8542 | -155.208 | 544 | 146.34 | 18:39:15.00 | N |
| O19K | 60.1952 | -154.3201 | 96 | 91.38 | 18:35:58.75 | Y |
| P17K | 59.1953 | -156.4394 | 91 | 245.40 | 18:44:18.00 | N |
| P18K | 59.3922 | -155.2292 | 680 | 190.90 | 18:41:27.00 | N |
| P19K | 59.6524 | -153.2319 | 629 | 122.03 | 18:37:30.40 | Y |
| RDDF | 60.5912 | -152.6883 | 1134 | 50.21 | 18:33:37.90 | Y |
| RDJH | 60.5905 | -152.8058 | 1414 | 48.37 | 18:33:32.35 | Y |
| RDSO | 60.4536 | -152.7453 | 1557 | 53.92 | 18:33:50.10 | Y |
| RDT | 60.5726 | -152.4079 | 945 | 55.22 | 18:33:52.56 | Y |
| RDWB | 60.486807 | -152.843634 | 1659 | 51.70 | 18:33:42.50 | Y |

SAND2026-18925O| | | | | | | |
|---|---|---|---|---|---|---|
| RED | 60.4196 | -152.7742 | 1071 | 55.87 | 18:33:56.43 | Y |
| REF | 60.4888 | -152.694 | 1641 | 52.74 | 18:33:44.15 | Y |
| SKN | 61.98 | -151.5317 | 581 | 168.07 | 18:39:42.00 | N |
| SPBG | 61.2591 | -152.3722 | 1087 | 82.82 | 18:35:26.60 | Y |
| SPCL | 61.1956 | -152.3399 | 1274 | 79.30 | 18:35:14.50 | Y |
| SPCN | 61.223675 | -152.18403 | 733 | 86.48 | 18:35:36.75 | Y |
| SPCP | 61.265465 | -152.155495 | 1640 | 89.92 | 18:35:48.70 | Y |
| SPNN | 61.3662 | -152.7012 | 1166 | 80.32 | 18:35:20.32 | Y |
| SPWE | 61.27346 | -152.559 | 1317 | 77.81 | 18:35:11.12 | Y |
| SSN | 61.4636 | -150.7467 | 1306 | 150.41 | 18:38:56.30 | Y |
| STLK | 61.4982 | -151.8349 | 945 | 118.23 | 18:37:19.00 | Y |
| WRH | 64.4716 | -148.0918 | 329 | 491.25 | 18:56:45.00 | N |

**Table S3.** Overview of infrasound array characteristics for arrays that recorded fireball-related signals. The infrasound signal period was measured at maximum amplitude using the 4 zero-crossings method (see Section 2.3 in the main text; Ens et al. (2012); ReVelle (1997); Silber (2024a)).

| FDSN Station | Station Latitude [deg] | Station Longitude [deg] | Station Altitude [m] | 3D distance to optical trajectory [km] | Signal onset time [UTC] | Ballistic arrival [Y/N] | Signal period amplitude [s] | Maximum amplitude [Pa] | Back-azimuth range [degrees] |
|---|---|---|---|---|---|---|---|---|---|
| BAEI | 61.132675 | -148.1219 | 592 | 229.84 | 18:47:30.00 | N | 0.838 | 0.152 | 257.35 - 268.98 |
| DLL | 59.139888 | -158.620929 | 57 | 337.87 | 18:49:30.00 | N | 1.295 | 0.082 | 51.14 - 59.49 |
| I53 | 64.875 | -147.86114 | 200 | 90.26 | 18:58:59.00 | N | 2.121 | 0.386 | 212.11 - 215.65 |
| KENI | 60.64137 | -151.0702 | 17 | 533.75 | 18:35:44.45 | Y | 1.564 | 1.208 | 241.68 - 283.56 |

Table S4. Results of the optical trajectory analysis.

|  | Beginning | End |
|---|---|---|
| **Time [UTC]** | 18:30:57.102681 | 18:31:02.749281 |
| **Latitude [°]** | 60.311735 | 60.909253 |
| **Longitude [°]** | -151.736025 | -153.728504 |
| **Height [km]** | 75.1947 | 31.2022 |
| **Velocity [km/s]** | 23.37 | 20.00 |
| **Azimuth [°]** | 302.253 | 302.253 |
| **Entry Angle [°]** | 18.99 | 18.99 |

Table S5. Orbital parameters of the fireball derived from the optical trajectory analysis.

| Parameter | Value | 1σ | Lower 95% Confidence Interval | Upper 95% Confidence Interval |
|---|---|---|---|---|
| Semi-major axis [AU] | 1.651308 | 0.000838 | 1.649631 | 1.652895 |
| Eccentricity | 0.655870 | 0.000140 | 0.655556 | 0.656110 |
| Inclination [°] | 10.719743 | 0.013193 | 10.684629 | 10.738264 |
| Argument of periapsis [°] | 264.378319 | 0.020424 | 264.345962 | 264.421849 |
| Longitude ascending node [°] | 214.449031 | 0.000029 | 214.448954 | 214.449072 |
| Perihelion [AU] | 0.568265 | 0.000112 | 0.568110 | 0.568526 |
| Aphelion | 2.734351 | 0.001605 | 2.731043 | 2.737347 |
| Tisserand's parameter | 3.987192 | 0.001488 | 3.984427 | 3.990229 |
| Last perihelion | 2460740.2306 | 0.0079 | 2460740.2140 | 2460740.2451 |
| Geocentric Right Ascension | 33.57659 | 0.00878 | 33.56199 | 33.59441 |
| Geocentric Declination | -2.69027 | 0.02045, | -2.72143 | -2.63767 |
| Geocentric velocity | 20.33532 | 0.00234 | 20.32956 | 20.33773 |